\documentclass[nonacm, sigconf]{acmart}

\usepackage[utf8]{inputenc} 
\usepackage[T1]{fontenc} 
\usepackage{graphicx} 
\usepackage{bibentry} 
\usepackage{float}
\usepackage{array}
\usepackage{booktabs}
\usepackage{multirow}
\usepackage{newfloat}
\usepackage{setspace}
\usepackage{tikz-cd}
\usepackage{lipsum}
\usepackage{wasysym}
\usepackage{tcolorbox}
\usepackage{rotating}
\usepackage{colortbl}
\usepackage{hhline}
\usepackage{caption}

\usepackage{tikz-cd}
\usepackage{lipsum}
\usepackage{wasysym}
\usepackage{tcolorbox}
\usepackage{hyperref}

\newcolumntype{L}{>{\centering\arraybackslash}m{6.2cm}}
\newcolumntype{G}{>{\centering\arraybackslash}m{5.5cm}}
\newcolumntype{A}{>{\centering\arraybackslash}m{2.4cm}}
\newcolumntype{B}{>{\centering\arraybackslash}m{7.5cm}}

\AtBeginDocument{%
  \providecommand\BibTeX{{%
    \normalfont B\kern-0.5em{\scshape i\kern-0.25em b}\kern-0.8em\TeX}}}

\setcopyright{acmcopyright}
\copyrightyear{2018}
\acmYear{2018}
\acmDOI{XXXXXXX.XXXXXXX}

\acmConference[Conference acronym 'XX]{Make sure to enter the correct
  conference title from your rights confirmation emai}{June 03--05,
  2018}{Woodstock, NY}
\acmPrice{15.00}
\acmISBN{978-1-4503-XXXX-X/18/06}

\begin{document}

\title{Avaliação dos impactos da decomposição de uma aplicação monolítica para microsserviços: Um estudo de caso}


\author{Tulio Ricardo Hoppen Barzotto}
\affiliation{%
  \institution{Universidade do Vale do Rio do Sinos (Unisinos)}
  \streetaddress{Av. Unisinos, 950 - Cristo Rei}
  \city{São Leopoldo, Rio Grande do Sul}
  \country{Brasil}}
\email{tulio.barzotto@gmail.com}

\author{Kleinner Farias}
\affiliation{%
  \institution{PPGCA, Universidade do Vale do Rio do Sinos (Unisinos)}
  \streetaddress{Av. Unisinos, 950 - Cristo Rei}
  \city{São Leopoldo, Rio Grande do Sul}
  \country{Brasil}}
\email{kleinnerfarias@unisinos.br}

\renewcommand{\shortauthors}{Barzotto and Farias}

\begin{abstract}
Aplicações monolíticas estão sendo decompostas para uma arquitetura de microsserviços, visando melhorar a manutenabilidade, perfomance e modularização. Embora tais decomposições tenham sido amplamente realizadas atualmente na indústria, pouco é reportado na literatura sobre os impactos destas decomposições. Este trabalho, portanto, reporta um estudo de caso realizado para investigar os impactos da decomposição de uma aplicação real da indústria para a arquitetura de microsserviços. A aplicação alvo do estudo refere-se a uma operação de saque, realizada por uma instituição financeira, a qual foi extraída de uma aplicação monolítica para uma aplicação baseada em microsserviços. Em particular, métricas foram aplicadas nas aplicação monolítica e na baseada em microsserviços, visando quantificar o acoplamento, coesão, consumo de CPU e consumo de memória. Os resultados obtidos apontam que a arquitetura de microsserviços gerou melhores resultados para as métricas de modularidade de software, além de menor consumo de memória e CPU. Por fim, este trabalho traz reflexão e aponta para desafios e direções futuras de pesquisa que precisam ser exploradas pela academia e a indústria.
\end{abstract}





\maketitle

\section{Introdução}
\label{ch:introducao}

As aplicações monolíticas podem ser caracterizadas como um único artefato de software executável, tendo seus módulos altamente acoplados e os requisitos implementados de forma entrelaçada e espalhada entre os módulos da aplicação~\cite{urdangarin2021mon4aware}. Atualmente, a arquitetura monolítica vem perdendo espaço para a arquitetura de microsserviços, evidenciando a popularidade, através de como as aplicações são entendidas, concebidas e desenhadas~\cite{dragoni2017microservices,rocha2018monolise}. Em particular, as constantes modificações dos requisitos e volatilidade dos ambientes de negócio provocam mudanças constantes das aplicações monolíticas~\cite{rubert2021effects,oliveira2018brcode}, dificultando as realizações das manutenções, aumentado as estimativas de esforço~\cite{carbonera2020software}, elevando o esforço cognitivo de desenvolvedores na compreensão de código~\cite{gonccales2021measuring}. Além disso, a ausência de documentação das aplicações monolíticas nas empresas aumenta o desafio da realização das manutenções~\cite{farias2018uml,junior2021survey}.

Com o uso da arquitetura de microsserviços, que divide a aplicação em um conjunto de serviços, que gera bases de códigos menores, que podem ser compreendidos mais facilmente pelos desenvolvedores, permite o desenvolvimento contínuo, sem afetar a aplicação como um todo, assim como uma série de outros benefícios, não encontrados na arquitetura monolítica. Nesta linha, é salientada a importância dos temas ``microsserviços'' e ``decomposição de aplicações monolíticas'', uma vez que possibilita ciclos de desenvolvimento de software menores, caracterizando aumento de performance e implantações mais ágeis, times menores e mais especializados. Destaca-se que a arquitetura monolítica possui uma forte desvantagem, já que, alterações na base de código ou implantações afetam a aplicação como um todo, seja por alterações de código ou disponibilidade durante novas implantações. Em contrapartida, a arquitetura de microsserviços possui uma estrutura modular que facilita as mudanças, tornando alterações de códigos e implantações pontuais, sem necessariamente afetar toda a aplicação ou necessidade de avisar todos os desenvolvedores sobre o novo código inserido.

Alguns trabalhos foram propostos, com finalidade de discussão do tema, por meio de análise comparativa. Autores como~\cite{dragoni2017microservices, knoche2018using} e outros, serviram como base para embasar o desenvolvimento do presente trabalho. Os trabalhos citados serviram especialmente para análise comparativa do tema, de acordo com critérios comparativos, como contexto de avaliação, domínio da aplicação, métodos de estudo empírico e tipos de métricas. 

Ressalta-se que a problemática do tema em questão, é escassa na literatura, visto que a arquitetura de microsserviços é relativamente recente, especialmente carece de estudos que propunham a decomposição de aplicações monolíticas, como base para a arquitetura de microsserviços. Além disso, costuma-se dar ênfase na arquitetura monolítica na maioria dos artigos e trabalhos sobre temas similares, com exceção de trabalhos como ~\cite{artigo1,artigo2,artigo3,artigo4,artigo5}, onde é realizada uma revisão sistemática, tanto da arquitetura monolítica, quanto na arquitetura de microsserviços e suas respectivas aplicações.

Neste sentido, essa pesquisa executa um estudo experimental, visando avaliar o impacto das métricas de consumo de CPU, consumo de memória e modularidade de software. Para isso foi executado um estudo empírico em que uma aplicação real da empresa hipotética \textit{Cooperativa Utile}, onde foi analisada a versão da aplicação monolítica em comparação com a sua versão equivalente, seguindo a arquitetura baseada em microsserviços. A modularidade de software foi analisada sob os atributos de acoplamento e coesão, usando 5 métricas voltadas para aplicações que utilizam linguagem orientada a objetos. As métricas de consumo de CPU e consumo de memória foram analisadas através de testes de carga, simulando a execução do fluxo de negócio equivalente em ambas versões da aplicação.

O presente trabalho se encontra estruturado da seguinte forma. Seção~\ref{ch:fundamentacao-teorica} apresenta a fundamentação teórica do tema dividido em subtópicos;~Seção~\ref{ch:trabalhos-relacionados} traz os trabalhos relacionados ao tema para análise comparativa, assim como a metodologia exposta de cada trabalho e subtópicos com a evidenciação das oportunidades do presente trabalho;~Seção~\ref{ch:metodologia} traz à luz, a metodologia utilizada no trabalho e subtópicos que salientam os objetivos e as questões de pesquisa explícitas, hipóteses, seleção da aplicação alvo, variáveis e métodos de quantificação, métricas selecionadas, procedimento de análise e experimental;~Seção~\ref{ch:resultados} traz os resultados obtidos; Por fim, a~Seção~\ref{ch:conclusao-trabalhos-futuros} aborda as conclusões e os trabalhos futuros. Já na estruturação pós-textual, encontram-se as referências bibliográficas utilizadas.

\section{Fundamentação teórica}
\label{ch:fundamentacao-teorica}

Esta seção aborda os conceitos teóricos usados durante a construção e desenvolvimento do estudo. A Seção está dividida da seguinte forma: a Seção~\ref{sec:arquitetura-monolitica} descreve os conceitos sobre a arquitetura monolítica; a Seção~\ref{sec:arquitetura-microsservicos} descreve os conceitos sobre a arquitetura de microsserviços; a Seção~\ref{sec:performance} descreve os conceitos sobre a performance de software; por fim a Seção~\ref{sec:modularizacao} descreve os conceitos sobre a modularização de software.

\subsection{Arquitetura monolítica}
\label{sec:arquitetura-monolitica}
Para um melhor entendimento sobre a arquitetura de microsserviços e como a tecnologia evoluiu para isto, será preciso primeiramente entender a arquitetura monolítica tradicional.

Na arquitetura monolítica, todas as funcionalidades estão encapsuladas em uma única aplicação, fazendo com que os módulos não possam ser executados independentemente ~\cite{dragoni2017microservices}. Este tipo de arquitetura torna a aplicação altamente acoplada e toda a lógica para processar a requisição é executada em um único processo, que usam os mesmos recursos de hardware, como memória \textit{RAM}, \textit{CPU} e armazenamento de dados. Devido ao fato de todo o desenvolvimento estar em um único executável, uma única alteração de código pode afetar todos os recursos que a aplicação provê, gerando necessariamente um novo \textit{build} e um \textit{redeploy} de toda a aplicação~\cite{richardson2018microservices}. 

Enquanto a arquitetura monolítica é uma boa escolha para iniciar um projeto, já que isso permite que você explore a complexidade de um sistema e os limites de seus componentes ~\cite{fowler_monolith_first}. Entretanto, os benefícios vão desaparecendo, conforme o código fonte da aplicação fica maior. Quanto maior o tamanho da aplicação, maior a complexidade, resultando em um grande número de dependências, o que causa alto acoplamento ~\cite{dragoni2017microservices, richardson2018microservices}. Com o objetivo de modernização destas aplicações, o processo de decomposição surge como atividade central, visando utilizar tecnologias emergentes e de arquiteturas de software distribuídas e de alta disponibilidade ~\cite{gysel2016service}.

\subsection{Arquitetura de microsserviços}
\label{sec:arquitetura-microsservicos}

De acordo com ~\cite{arun}, os microsserviços são resultados de uma abordagem arquitetônica, focada na decomposição de aplicações em serviços, com propósito único e com baixo acoplamento, sendo gerenciadas por equipes multifuncionais, para entrega e manutenção de sistemas de software complexos rapidamente. Para ~\cite{fowler_microservices}, a arquitetura de microsserviços consiste em uma abordagem para desenvolver um único aplicativo, como um conjunto de pequenos serviços, cada um sendo executado de forma isolada e se comunicando de forma leve, geralmente uma \textit{API} de recursos \textit{HTTP}.

Segundo~\cite{fowler_microservices}, não existe em particular uma definição do que seja arquitetura de microsserviços, mas existem certas características que o tornam sujeitos a classificar como arquitetura de microsserviços, "Como acontece com qualquer definição que descreve características comuns, nem todas as arquiteturas de microsserviços têm todas as características, mas esperamos que a maioria das arquiteturas de microsserviços exibam a maioria das características"~\cite[Traduzido pelo autor]{fowler_microservices}. Para~\cite{newman2015building}, essas características são mais como princípios da arquitetura de microsserviços, os quais são definidos como:

\begin{enumerate}
    \item ``Modelar em torno de conceitos de negócios'', para serem representados como contextos limitados e modelos de domínio de acordo com padrões do \textit{Domain-Driven Design} (DDD)~\cite{evans2009domain}.
    \item ``Adotar uma cultura de automação'' em testes e implantação; praticar entrega contínua.
    \item ``Ocultar detalhes da implementação interna'', como bancos de dados; definir Interfaces de programação de aplicativos (APIs) independentes de tecnologia.
    \item ``Descentralizar todas as coisas'': por exemplo, aplicar a governança compartilhada, prefira o serviço de coreografia ao invés de orquestração, use \textit{middleware} burro, mas \textit{endpoints} inteligentes.
    \item Tornar os serviços ``implantáveis de forma independente'', por exemplo, deixar versionado (serviço) \textit{endpoints} coexistem; implantar apenas um serviço por \textit{host} (virtual).
    \item ``Isolar falha'', por exemplo, introduz disjuntores para tornar os serviços robustos.
    \item Ser ``altamente observável'', por exemplo, por meio de monitoramento semântico com dados de agregação.

\end{enumerate}

Para fins de conceituação dos diferentes autores, evidencia-se que ao contrário das nove características que~\cite{fowler_microservices} expõem, para~\cite{newman2015building}, apenas sete princípios são válidos para definir a arquitetura de microsserviços. Porém, tanto os princípios de~\cite{newman2015building} quanto às características de ~\cite{fowler_microservices}, mesclam-se, seja através da modelagem da aplicação em torno do negócio, como a descentralização da governança.

\subsection{Performance}
\label{sec:performance}
Segundo ~\cite{rafighi2015studying}, a performance é um fator crucial, pois ela impacta diretamente na experiência do usuário ao utilizar um determinado sistema. A performance está invariavelmente atrelada a capacidade da máquina e sua composição arquitetural, podendo ser medida em vários parâmetros, como taxa de transferência, latência e largura de banda do sistema~\cite{desai2016survey}. Para fins de catalogação de medição da performance, é inerente que o desempenho da \textit{CPU}, sendo a taxa de transferência utilizada como quesito para medir a saída de carga de trabalho, o desempenho da memória que tem como parâmetro a medição da largura da banda na velocidade de acesso à memória e operações, assim como o desempenho da rede, de disco e outros, os quais estão ligados à performance, tanto da arquitetura, quanto da máquina, sendo usados como atributos para conceituação ~\cite{desai2016survey}.

De acordo com ~\cite{rocha2018monolise}, são os recursos utilizados sob condições estabelecidas que representam o desempenho/performance da arquitetura. Esses mesmos recursos estão ligados à eficiência do desempenho em características, como comportamento do tempo (tempo de resposta), utilização de recursos (tipos de recursos utilizados por um produto), dentre outros. Também são atributos usados para medição de performance, ou ao menos caracterização e conectividade deste desempenho, a compatibilidade, usabilidade, confiabilidade, segurança, manutenibilidade e portabilidade~\cite{rocha2018monolise}. A arquitetura de microsserviços traz complexidade aos testes de desempenho, que são classificados como: tipo caixa preta, portanto, o tipo de teste mais compatível é o \textit{end-to-end}~\cite{camargo2016abordagem}. Segundo os preceitos de ~\cite{thakare2012software}, são os testes de desempenho que têm como finalidade a verificação do software e o cumprimento dos requisitos pré-estabelecidos, como o tempo de resposta, vazão e disponibilidade.

\subsection{Modularização}
\label{sec:modularizacao}
A conceituação de modularização se dá através do compartilhamento de que é uma atividade na qual a estruturação em módulos é adotada, portanto, um sistema complexo é estruturado em vários subsistemas independentes (módulos) ~\cite{ramos2016analise}. Ressalta-se a importância da diferenciação entre módulo (relacionado a uma unidade funcional independente em relação ao propósito do produto), modularização (estruturação em vários subsistemas) e modularidade (concepção de produtos complexos a partir da combinação de módulos relativamente simples) ~\cite{ramos2016analise}. Para~\cite{ulrich1994fundamentals}, a modularização se dá através de duas características inerentes ao conceito: 1) similaridade entre a arquitetura física e funcional do produto; 2) minimização do grau de interação entre os componentes físicos.

A partir do conceito de modularização, surge o termo \textit{bad smells}, que segundo~\cite{fowler1997refactoring} \textit{smells} são estruturas no código que sugerem a possibilidade de refatoração. Já ~\cite{palomba2013detecting}, conceitua \textit{smells} como \textit{"symptoms of poor design and implementation choices"}, sendo descrito na literatura, uma catalogação de 104 \textit{smells}, sendo os mais importantes \textit{duplicate code} (DC), \textit{large class} (LC), \textit{feature envy} (FE) e outros. Alguns não apresentam técnicas, ou até mesmo ferramentas para identificação de suas instâncias, tornando assim o uso de estratégias, uma opção para identificação de alguns \textit{bad smells}, portanto, basicamente os \textit{bad smells}, descrevem possíveis problemas em determinado código, possibilitando oportunidades de refatoração~~\cite{ramos2016analise}. A evidenciação de \textit{bad smells}, reforça a necessidade de exposição da divida técnica.

A dívida técnica é nada mais do que reflexão dos compromissos técnicos, que podem resultar em benefícios a curto prazo, mas em contrapartida, podem causar danos e prejuízos à qualidade de um sistema de software a longo prazo~\cite{brown2010managing}. Segundo ressalta~\cite{allman2012managing}, a dívida técnica é inevitável, sendo assim, o enfoque principal não é tentar eliminá-la, mas mantê-la sob controle, por meio de seu gerenciamento. É por conta disso que, a dívida técnica é causada a partir de uma tomada de decisão, ou um processo, uma ação, ou falta dela, resultando pela pressão de uma programação, ou de cronograma, pela indisponibilidade de uma pessoa chave, ou pela falta de informações sobre um recurso técnico~\cite{avgeriou2016managing}.

\section{Trabalhos relacionados}
\label{ch:trabalhos-relacionados}
Esta Seção realiza uma análise comparativa dos trabalhos relacionados. A análise tem por objetivo identificar, destacar as características comuns e as diferenças entre os estudos já realizados e o trabalho proposto. A seção está divido da seguinte forma: a Seção~\ref{sec:trab-relac-metodologia} descreve a metodologia utilizada para escolha dos trabalhos relacionados; a Seção~\ref{sec:trab-relac-analise} é realizada a análise de cinco artigos que satisfazem os critérios de seleção; a Seção~\ref{sec:trab-relac-analise-comparativa} realiza a comparação dos trabalhos, mediante critérios definidos; por último as oportunidades de pesquisa identificadas.

\subsection{Metodologia para escolha dos trabalhos}
\label{sec:trab-relac-metodologia}
Este trabalho utilizou como base de dados o \textit{Google Scholar}. Foi utilizada a \textit{query} de pesquisa com as seguintes palavras-chave:

("\textit{microservice*}" \textbf{OR} "\textit{micro-service}") \textbf{AND} ("\textit{monolith*}") \textbf{AND} ("\textit{comparative study}" \textbf{OR} "\textit{empirical study}" \textbf{OR} "\textit{performance}")

Com base nos resultados obtidos na busca, foram selecionados cinco artigos pela similaridade com o tema em estudo.

\subsection{Análise dos trabalhos relacionados}
\label{sec:trab-relac-analise}
Nesta seção, será realizada uma análise comparativa de cinco trabalhos que abordam o tema de análise de performance entre aplicações com arquitetura monolítica e microsserviços.

\textbf{Gos and Zabierowski (2020)~\cite{artigo1}}. Neste estudo, os autores propõem uma análise comparativa, entre uma aplicação desenvolvida em uma arquitetura monolítica e outra equivalente desenvolvida em uma arquitetura baseada em microsserviços. A aplicação alvo é um sistema de aluguel de carros, contendo as funcionalidades de consulta, cadastro, aluguel e atualização do status. A aplicação monolítica foi desenvolvida em \textit{Java} utilizando \textit{Spring} e um único banco de dados \textit{PostgreSQL}. A aplicação baseada em arquitetura de microsserviços também foi desenvolvida em \textit{Java} utilizando \textit{Spring}, sendo composta por quatro serviços autônomos e independentes. Os testes de carga foram realizados utilizando a ferramenta \textit{Gatling}, onde foram separados em dois cenários, o primeiro visa simular 1000 requisições feitas por 30 usuários simultaneamente, já no segundo cenário é um total de 10000 requisições feitas por 30 usuários simultaneamente. Ambos cenários foram aplicados em requisições \textit{GET} e requisições \textit{POST}. Os autores concluem que a arquitetura baseada em microsserviços é mais eficiente em lidar com um número grande de requisições simultâneas, além de permitir construir um software de alta qualidade, fácil de escalar, confiável e a longo prazo mais fácil de manter. Também consideram que a arquitetura monolítica é mais eficiente em cargas mais baixas e mais fácil de ser desenvolvida.

\textbf{Villamizar \textit{et al.} (2015)~\cite{artigo2}}. Neste estudo, os autores buscam avaliar em um cenário real as implicações de uso de uma arquitetura baseada em microsserviços. A aplicação alvo é um sistema de gestão de empréstimos realizados por uma instituição financeira aos seus clientes. O estudo selecionou apenas dois dos serviços mais utilizados, o primeiro serviço intitulado \textit{$S^1$} é responsável por gerar um plano de pagamento, contendo as parcelas do empréstimo contratado, o segundo serviço \textit{$S^2$} é responsável por retornar um plano de pagamento existente com o respectivo conjunto de parcelas. O \textit{$S^1$} utiliza algorítimos para gerar cada plano de pagamento, usando excessivamente recursos da \textit{CPU}, o \textit{$S^2$} realiza a consulta do plano de pagamento através de um identificador único na base de dados. Considerando a aplicação desenvolvida na arquitetura monolítica onde ambos serviços estão implementados na mesma aplicação e expostos via \textit{endpoints}, temos os seguintes tempos médios de respostas, \textit{$S^1$} com 3.000 milissegundos e \textit{$S^2$} em torno de 300 milissegundos. Na versão da aplicação desenvolvida utilizando arquitetura baseada em microsserviços, foram criados dois microsserviços, um para implementar as regras do \textit{$S^1$} e outro as regras do \textit{$S^2$}. Ambas abordagens da aplicação foram desenvolvidas com Java, utilizando o \textit{framework Play} e hospedados em uma solução \textit{cloud} da \textit{Amazon Web Services (AWS)}. Utilizaram o \textit{JMeter} para realizar os testes de carga e avaliar a performance, onde foram criados dois cenários de teste aplicados, nas duas aplicações, sendo que o primeiro cenário de teste é focado no \textit{$S^1$} executando 30 requisições por minuto, no segundo cenário de teste focado no \textit{$S^2$} foram executados 1.100 requisições por minuto. Os autores concluem que os microsserviços não impactam consideravelmente na latência de resposta usando mais instâncias, no entanto, a granularidade dos microsserviços permite escalonar pontos específicos da aplicação, reduzindo os custos de infraestrutura. Também consideram que há benefício no desenvolvimento de software utilizando arquitetura baseada em microsserviços, permitindo com que pequenas equipes trabalhem em pequenos microsserviços de forma independente. Os autores identificaram desafios de sistemas distribuídos ao introduzirem a arquitetura baseada em microsserviços, os quais são gerenciados de maneira mais simples, em uma arquitetura monolítica.

\textbf{Rudrabhatla (2020) \cite{artigo3}}. Este estudo aborda três técnicas de decomposição de aplicações monolíticas para microsserviços e realiza uma comparação de performance e latência entre elas. A aplicação alvo é um módulo de uma aplicação \textit{e-commerce}, composta inicialmente por três entidades de negócios. Na primeira técnica de decomposição abordada, sugere o uso da técnica \textit{Domain-Driven Design}, onde usando \textit{Common Closure Principle} são identificadas as classes que são impactadas pela mesma regra de negócio, devem estar no mesmo pacote em um único microsserviço. No sentido de que cada microsserviço seja autônomo, cada um possui uma base dados própria e isolada dos demais. A segunda técnica aborda a decomposição baseada em \textit{Business capability} onde as principais entidades são normalizadas e os microsserviços são construídos considerando a transação de negócio. A granularidade das entidades são definidas até o ponto em que possam ser consideradas autônomas e operações de \textit{CRUD} são criadas para gerenciamento. Na terceira e última técnica abordada, a decomposição é baseada em uma abordagem híbrida, combinando as duas primeiras técnicas, onde são identificados subdomínios utilizando \textit{Business capability} gerando microsserviços unificados. O autor realiza uma comparação de desempenho e latência entre as três abordagens, observando os tempos de respostas para persistência dos dados em banco de dados. Como resultado da análise comparativa, o autor pode demonstrar que a técnica \textit{Domain-Driven Design} foi superior a \textit{Business capability}, mas que em uma aplicação \textit{real time} a abordagem híbrida produziu melhores resultados.

\textbf{Tapia \textit{et al.} (2020) \cite{artigo4}}. Este estudo propõe uma análise comparativa de performance entre arquitetura monolítica e arquitetura baseada em microsserviços. A aplicação alvo de estudo é um fórum, contendo três entidades de negócio. Primeiramente são descritas as estruturas técnicas, onde aplicação monolítica é desenvolvida utilizando a linguagem \textit{Node.js}, contendo as entidades de negócio implementadas na mesma aplicação, hospedada em uma máquina virtual. A aplicação equivalente em arquitetura baseada em microsserviços também é desenvolvida utilizando a linguagem \textit{Node.js}. As entidades de negócios estão separadas em aplicações distintas, conteinerizadas, utilizando \textit{Docker} e hospedadas na estrutura de \textit{cloud} da \textit{Amazon}. O processo de coleta de dados é feito através de testes de estresse, separados por dois cenários, onde o primeiro realiza requisições via \textit{HTTP} para gerar dados no banco de dados, o segundo cenário realiza requisições via \textit{HTTP} para selecionar dados persistidos no banco de dados. O estudo também aplica o modelo matemático \textit{Non-Parametric Regression Model} para relacionar dados de recursos computacionais utilizados e assim poder avaliar os dados coletados. Os autores concluem que o uso da arquitetura baseada em microsserviços é mais vantajosa em relação aos recursos de hardware e redução de custos.

\textbf{Bjørndal \textit{et al.} (2020) \cite{artigo5}}. Neste estudo os autores procuram verificar os benefícios de migração de uma arquitetura monolítica para uma arquitetura baseada em microsserviços, realizando experimentos de \textit{benchmarking}. A aplicação alvo é um sistema fictício de biblioteca, contendo as funcionalidades necessárias para login e empréstimo de livros. A aplicação com arquitetura monolítica foi desenvolvida utilizando \textit{ASP.NET Core} e base de dados centralizada utilizando \textit{SQL Server}, já a aplicação com arquitetura baseada em microsserviços gerou quatro serviços, representando as entidades de domínio, cada um com seu contexto limitado e base de dados própria. Foram aplicados dois experimentos de \textit{benchmarking}, onde a aplicação foi executada em uma estrutura local e outra hospedada no \textit{Azure Cloud}. As métricas utilizadas em ambos experimentos foram \textit{Throughput}, \textit{Latency}, \textit{Scalability} e recursos de hardware (\textit{CPU}, \textit{Memory}, \textit{Network}). Os experimentos são compostos por uma execução simples e outra complexa, todos eles executados utilizando \textit{JMeter}. Os autores concluem que a arquitetura monolítica foi melhor em todas métricas, exceto na escalabilidade, entretanto consideram pontos em que o ambiente e aplicação podem ter levado a este resultado, como o tamanho reduzido da aplicação, se comparado a uma aplicação real, assim como gargalos de uso entre \textit{kubernetes}, \textit{Docker} e os próprios bancos de dados. Concluem também que apesar das considerações anteriores, a arquitetura baseada em microsserviços se torna uma melhor alternativa, considerando uma grande quantidade de usuários simultâneos, devido a métrica de escalabilidade ter sido melhor.

\subsection{Análise comparativa dos trabalhos relacionados}
\label{sec:trab-relac-analise-comparativa}
Nesta seção será realizada a comparação dos trabalhos selecionados no item 3.2 com base nos critérios comparativos definidos:

\begin{itemize}
    \item \textbf{Contexto de Avaliação:} Este critério irá avaliar se a proposta do trabalho foi avaliada em um ambiente \textit{acadêmico} ou na \textit{indústria}.
    \item \textbf{Domínio da aplicação:} Este critério irá avaliar qual o domínio da aplicação alvo: \textit{automotivo}, \textit{comunicação}, \textit{financeiro}, \textit{literatura}, \textit{varejo}.
    \item \textbf{Métodos de estudo empírico:} Este critério irá avaliar qual o método de avaliação foi utilizado para avaliar as arquiteturas propostas pelos trabalhos: \textit{estudo de caso}, \textit{experimento} ou \textit{levantamento}.
    \item \textbf{Tipos de Métricas:} Este critério irá avaliar quais métricas de avaliação foram utilizadas pelos trabalhos: \textit{custos}, \textit{disponibilidade}, \textit{escalabilidade}, \textit{performance}, \textit{qualidade de código}, \textit{segurança}.
\end{itemize}

\begin{table*}
\centering
\caption{Comparação dos trabalhos relacionados}
\arrayrulecolor{black}
\begin{tabular}{|l|l|c|c|c|c|c|} 
\hline
\multicolumn{2}{|c|}{\multirow{2}{*}{\textbf{Critérios }}}                                                                                                               & \multicolumn{5}{c|}{\textbf{Trabalhos }}                                                                                                                                                                                                                                                                                                                                                                                                                                                                                                                                                                                                                                          \\ 
\cline{3-7}
\multicolumn{2}{|c|}{}                                                                                                                                                   & \multicolumn{1}{l|}{\begin{sideways}Gos and Zabierowski (2020)~\cite{artigo1}\end{sideways}} & \multicolumn{1}{l|}{\begin{sideways}Villamizar \textit{et al}. (2015)~\cite{artigo2}\end{sideways}} & \multicolumn{1}{l|}{\begin{sideways}Rudrabhatla (2020)~\cite{artigo3}\end{sideways}} & \multicolumn{1}{l|}{\begin{sideways}Tapia \textit{et al}. (2020)~\cite{artigo4}\end{sideways}} & \multicolumn{1}{l|}{\begin{sideways}Bjørndal \textit{et al.} (2020)~\cite{artigo5}\end{sideways}}  \\ 
\hline
\multirow{2}{*}{Contexto de avaliação}                                                                             & Acadêmico                                           & +                                                                                                             & -                                                                                                                                                              & +                                                                                                                                   & +                                                                                                                                & +                                                                                                                        \\ 
\hhline{|~------|}
                                                                                                                   & {\cellcolor[rgb]{0.788,0.788,0.788}}Indústria       & {\cellcolor[rgb]{0.788,0.788,0.788}}-                                                                         & {\cellcolor[rgb]{0.788,0.788,0.788}}+                                                                                                                         & {\cellcolor[rgb]{0.788,0.788,0.788}}-                                                                                               & {\cellcolor[rgb]{0.788,0.788,0.788}}-                                                                                            & {\cellcolor[rgb]{0.788,0.788,0.788}}-                                                                                    \\ 
\hline
\multirow{5}{*}{Domínio da aplicação}                                                                              & Automotivo                                          & +                                                                                                             & -                                                                                                                                                             & -                                                                                                                                   & -                                                                                                                                & -                                                                                                                        \\ 
\hhline{|~------|}
                                                                                                                   & {\cellcolor[rgb]{0.788,0.788,0.788}}Comunicação     & {\cellcolor[rgb]{0.788,0.788,0.788}}-                                                                         & {\cellcolor[rgb]{0.788,0.788,0.788}}-                                                                                                                         & {\cellcolor[rgb]{0.788,0.788,0.788}}-                                                                                               & {\cellcolor[rgb]{0.788,0.788,0.788}}+                                                                                            & {\cellcolor[rgb]{0.788,0.788,0.788}}-                                                                                    \\ 
\cline{2-7}
                                                                                                                   & Financeiro                                             & -                                                                                                             & +                                                                                                                                                             & -                                                                                                                                   & -                                                                                                                                & -                                                                                                                        \\ 
\hhline{|~------|}
                                                                                                                   & {\cellcolor[rgb]{0.788,0.788,0.788}}Literatura      & {\cellcolor[rgb]{0.788,0.788,0.788}}-                                                                         & {\cellcolor[rgb]{0.788,0.788,0.788}}-                                                                                                                         & {\cellcolor[rgb]{0.788,0.788,0.788}}-                                                                                               & {\cellcolor[rgb]{0.788,0.788,0.788}}-                                                                                            & {\cellcolor[rgb]{0.788,0.788,0.788}}+                                                                                    \\ 
\cline{2-7}
                                                                                                                   & Varejo                                              & -                                                                                                             &                                                                                                                                                               & +                                                                                                                                   & -                                                                                                                                & -                                                                                                                        \\ 
\hline
\rowcolor[rgb]{0.788,0.788,0.788} {\cellcolor[rgb]{0.788,0.788,0.788}}                                             & Estudo de caso                                      & +                                                                                                             & +                                                                                                                                                             & +                                                                                                                                   & +                                                                                                                                & +                                                                                                                        \\ 
\hhline{|>{\arrayrulecolor[rgb]{0.788,0.788,0.788}}->{\arrayrulecolor{black}}------|}
{\cellcolor[rgb]{0.788,0.788,0.788}}                                                                               & Experimento controlado                              & -                                                                                                             & -                                                                                                                                                             & -                                                                                                                                   & -                                                                                                                                & -                                                                                                                        \\ 
\hhline{|>{\arrayrulecolor[rgb]{0.788,0.788,0.788}}->{\arrayrulecolor{black}}------|}
\rowcolor[rgb]{0.788,0.788,0.788} \multirow{-3}{*}{{\cellcolor[rgb]{0.788,0.788,0.788}}Métodos de estudo empírico} & Survey (Levantamento)                               & -                                                                                                             & -                                                                                                                                                             & -                                                                                                                                   & -                                                                                                                                & -                                                                                                                        \\ 
\hline
\multirow{6}{*}{Tipo  de  Métricas}                                                                                & Custos                                              & -                                                                                                             & +                                                                                                                                                             & -                                                                                                                                   & -                                                                                                                                & -                                                                                                                        \\ 
\hhline{|~------|}
                                                                                                                   & {\cellcolor[rgb]{0.788,0.788,0.788}}Disponibilidade & {\cellcolor[rgb]{0.788,0.788,0.788}}-                                                                         & {\cellcolor[rgb]{0.788,0.788,0.788}}-                                                                                                                         & {\cellcolor[rgb]{0.788,0.788,0.788}}-                                                                                               & {\cellcolor[rgb]{0.788,0.788,0.788}}-                                                                                            & {\cellcolor[rgb]{0.788,0.788,0.788}}-                                                                                    \\ 
\cline{2-7}
                                                                                                                   & Escalabilidade                                      & -                                                                                                             & +                                                                                                                                                             & -                                                                                                                                   & +                                                                                                                                & +                                                                                                                        \\ 
\hhline{|~------|}
                                                                                                                   & {\cellcolor[rgb]{0.788,0.788,0.788}}Performance     & {\cellcolor[rgb]{0.788,0.788,0.788}}+                                                                         & {\cellcolor[rgb]{0.788,0.788,0.788}}+                                                                                                                         & {\cellcolor[rgb]{0.788,0.788,0.788}}+                                                                                               & {\cellcolor[rgb]{0.788,0.788,0.788}}+                                                                                            & {\cellcolor[rgb]{0.788,0.788,0.788}}+                                                                                    \\ 
\cline{2-7}
                                                                                                                   & Qualidade de código                                 & -                                                                                                             & -                                                                                                                                                             & -                                                                                                                                   & -                                                                                                                                & -                                                                                                                        \\ 
\hhline{|~------|}
                                                                                                                   & {\cellcolor[rgb]{0.788,0.788,0.788}}Segurança       & {\cellcolor[rgb]{0.788,0.788,0.788}}-                                                                         & {\cellcolor[rgb]{0.788,0.788,0.788}}-                                                                                                                         & {\cellcolor[rgb]{0.788,0.788,0.788}}-                                                                                               & {\cellcolor[rgb]{0.788,0.788,0.788}}-                                                                                            & {\cellcolor[rgb]{0.788,0.788,0.788}}-                                                                                    \\
\hline
\end{tabular}
\caption*{Legenda: (+) Aplicado (-) Não aplicado}
\end{table*}

Sobre as principais contribuições, apenas o trabalho~\cite{artigo2} foi aplicado na indústria. Os trabalhos~\cite{artigo1}, ~\cite{artigo3}, ~\cite{artigo4} e~\cite{artigo5} foram aplicados na acadêmia. O domínio da aplicação alvo no estudo~\cite{artigo1} foi automotivo, no estudo ~\cite{artigo2} foi financeiro, no estudo ~\cite{artigo3} foi varejo, no estudo ~\cite{artigo4} foi comunicação e no estudo~\cite{artigo5} foi literatura. Como métodos de avaliação dos trabalhos selecionados, todos os trabalhos apresentaram um estudo de caso de suas proposta. Dos estudos selecionados apenas o~\cite{artigo2} considerou os custos como métrica. Os estudos~\cite{artigo4} e~\cite{artigo5} avaliaram como métrica a escalabilidade. No entanto, todos os estudos consideraram a métrica de performance para avaliação.

\textbf{Oportunidades de pesquisa.} Após a análise dos artigos, as seguintes oportunidades de pesquisa foram identificadas: (1) avaliação entre ambas arquiteturas considerando a performance; e (2) avaliação da complexidade de código considerando métricas de modularidade de software;

Destas oportunidades, este trabalho procura gerar conhecimento empírico sobre os impactos da arquitetura monolítica e da arquitetura baseada em microsserviços na modularidade e na performance.

\section{Metodologia}
\label{ch:metodologia}
Esta seção apresenta as principais decisões que fundamentam a análise comparativa do estudo de caso~\cite{garcia2006modularizing,Baumann09}. Para começar, são apresentados os objetivos e as questões de pesquisa (Seção~\ref{sec:objetivo-questao-pesquisa}). Em seguida, são descritas as hipóteses (Seção~\ref{sec:hipoteses}). Na sequência a seleção da aplicação alvo (Seção~\ref{sec:selecao-aplicacao-alvo}). As variáveis e métodos de quantificação considerados, também são discutidos em detalhes (Seção~\ref{sec:variaveis-metodo-quantificacao}). A seleção de métricas é abordada na sequência (Seção~\ref{sec:metricas-selecionadas}). A descrição do procedimento de análise (Seção~\ref{sec:procedimento-analise}). Finalmente, o procedimento experimental é apresentado (Seção~\ref{sec:procedimento-experimental}). Todas essas etapas metodológicas foram baseadas em diretrizes práticas em estudos empíricos~\cite{farias2015evaluating}.

\subsection{Objetivo e Questões de Pesquisa}
\label{sec:objetivo-questao-pesquisa}

Este estudo tenta essencialmente avaliar os efeitos do estilo de arquitetura de microsserviços na performance e modularidade de software. Busca-se investigar os efeitos da performance de hardware, através de duas métricas: consumo de memória e consumo de CPU~\cite{artigo5}. No entanto, busca-se investigar os efeitos na modularidade de software através de duas variáveis~\cite{chidamber1994metrics}: acoplamento e coesão. Esses efeitos são investigados no contexto de uma aplicação real do sistema financeiro (Seção~\ref{sec:selecao-aplicacao-alvo}), a qual foi construída seguindo o estilo arquitetural de microsserviços, resultado de um processo de decomposição de uma aplicação monolítica (Seção~\ref{sec:decomposicao-aplicao-alvo}). Com isso em mente, o objetivo deste estudo é estabelecido com base no modelo GQM~\cite{sjoberg2002conducting}.

\begin{flushleft}
    \hspace{1cm} \textbf{analisar} estilos arquiteturais\\
    \hspace{1cm} \textbf{com o propósito de} investigar seus efeitos\\
    \hspace{1cm} \textbf{com relação à} modularidade de software e performance\\
    \hspace{1cm} \textbf{da perspectiva dos} desenvolvedores\\
    \hspace{1cm} \textbf{no contexto da} evolução tecnológica\\
\end{flushleft}

Este estudo tem como foco principalmente avaliar os efeitos da decomposição de uma aplicação monolítica em uma arquitetura de microsserviços. Desta forma, o foco será em duas questões de pesquisa:

\begin{itemize}
    \item \textbf{QP1:} Qual seria o efeito na \textit{modularidade de software} de aplicações monolíticas em comparação com aplicações baseadas em microsserviços?
    \item \textbf{QP2:} Qual seria o efeito na \textit{performance} de aplicações monolíticas em comparação com aplicações baseadas em microsserviços?
\end{itemize}

\subsection{Hipóteses}
\label{sec:hipoteses}

\textit{Hipótese 1.} Conjectura-se que a arquitetura monolítica tende a gerar aplicações maiores, com um grande número de dependências, o que causa alto acoplamento do software~\cite{dragoni2017microservices, richardson2018microservices}. No entanto, o estilo arquitetural de microsserviços separa a aplicação em serviços independentes~\cite{evans2009domain}, promovendo o baixo acoplamento e a alta coesão~\cite{newman2015building}. Consequentemente, conjectura-se que os microsserviços resultantes da decomposição irão possuir maior modularização do software. Desta forma,  declarando as hipóteses nula e alternativa da seguinte forma:

\begin{center}
	\fbox{%
		\parbox{0.75\linewidth}{
			
			\noindent \textbf{Hipótese Nula 1, H\textsubscript{1--0}:} Não há diferença entre as métricas de modularidade de software utilizando arquitetura de microsserviços em comparação com a arquitetura monolítica.
			\begin{equation}
			H\textsubscript{1--0}:\resizebox{.8\hsize}{!}{$Modularidade(Microsservico) = Modularidade(Monolitica)$}
			\end{equation}
			\noindent \textbf{Hipótese alternativa 1, H\textsubscript{1--1}:} A arquitetura de microsserviços possui maior modularidade de software em comparação com a arquitetura monolítica.
			\begin{equation}
			H\textsubscript{1--1}:\resizebox{.8\hsize}{!}{$Modularidade(Microsservico) > Modularidade(Monolitica)$}
			\end{equation}
		}
	}
\end{center}

Testando esta primeira hipótese, será confirmado (ou não) o aumento da modularidade de software, através de métricas avaliando cada aspecto do software gerado.\\

\textit{Hipótese 2.} Conforme mencionado anteriormente, a arquitetura de microsserviços gera serviços independentes entre si, comunicando-se de maneira leve, com recursos de hardware próprios. Entretanto, a arquitetura monolítica segue uma abstração dos componentes que depende do compartilhamento de recursos do mesmo servidor onde os componente não são executáveis de forma independente~\cite{newman2015building}. No entanto, os ganhos de performance provenientes do uso de cada arquitetura não são evidentes~\cite{artigo5}. Consequentemente, conjectura-se que a arquitetura baseada em microsserviços exigirá um menor consumo de memória e menor uso de CPU. No entanto, não é de forma alguma óbvio que esta hipótese seja válida. Talvez, a quantidade de requisições simultâneas gere um consumo menor de memória e CPU entre os microsserviços; ou ambas arquiteturas obtenham resultados semelhantes. Com base nesta declaração, declaro as hipóteses nula e alternativa da seguinte forma:

\begin{center}
	\fbox{%
		\parbox{0.75\linewidth}{
			
			\noindent \textbf{Hipótese Nula 2, H\textsubscript{2--0}:} Não há diferença no uso dos recursos de hardware utilizando arquitetura de microsserviços em comparação com a arquitetura monolítica.
			\begin{equation}
			H\textsubscript{2--0}:\resizebox{.8\hsize}{!}{$ConsumoCPU(Microsservico) = ConsumoCPU(Monolitica)$}
			\end{equation}
			\begin{equation}
			H\textsubscript{2--0}:\resizebox{.8\hsize}{!}{$ConsumoMemoria(Microsservico) = ConsumoMemoria(Monolitica)$}
			\end{equation}
			\noindent \textbf{Hipótese alternativa 2, H\textsubscript{2--1}:} Há uma redução do consumo de memória e CPU utilizando arquitetura de microsserviços em comparação com a arquitetura monolítica.
			\begin{equation}
			H\textsubscript{2--1}:\resizebox{.8\hsize}{!}{$ConsumoCPU(Microsservico) < ConsumoCPU(Monolitica)$}
			\end{equation}
			\begin{equation}
			H\textsubscript{2--1}:\resizebox{.8\hsize}{!}{$ConsumoMemoria(Microsservico) < ConsumoMemoria(Monolitica)$}
			\end{equation}
		}
	}
\end{center}

Testando esta segunda hipótese, será possível avaliar o uso dos recursos de hardware, comparando os recursos utilizados em ambas abordagens arquiteturais.\\

\subsection{Seleção da Aplicação Alvo}
\label{sec:selecao-aplicacao-alvo}
Para a execução do estudo de caso, foram selecionados projetos no repositório de código fonte da \textit{Cooperativa Utile}. A coleta dos projeto considerou algumas características:
\begin{enumerate}
    \item Aplicação ser caracterizada como monolítica;
    \item Aplicação escrita utilizando linguagem Java;
    \item Decomposição da aplicação estar finalizada ou em etapa de finalização;
\end{enumerate}
Após a conclusão da busca dos projetos, obteve-se uma amostra inicial de 3 projetos. Os 3 projetos foram avaliados, tanto em qual etapa de entrega estava quanto sua complexidade. Os projetos em etapa de execução intermediária ou análise da decomposição foram descartados. A aplicação selecionada foi projetada para atender diversos recursos dos terminais de autoatendimento aos clientes da instituição financeira, o objeto de estudo será focado no módulo de saque em dinheiro.

Esta aplicação atualmente atende diversos tipos de terminais de autoatendimento, alguns dos recursos são: extrato de conta corrente, extrato e antecipação de cota capital, seguro veicular, transferência eletrônica, depósito de cheque, pagamento de DOC, entre outras funcionalidades.

\begin{figure}[H]
  \centering
  \includegraphics[width=0.35\textwidth]{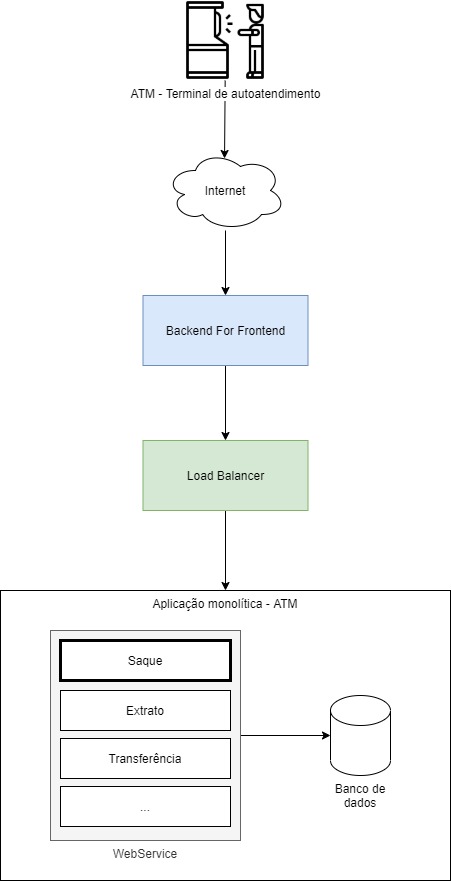}
  \caption{Diagrama geral da arquitetura monolítica}
  \label{fig:diagrama-arq-monolitica}
\end{figure}

\subsubsection{Decomposição}
\label{sec:decomposicao-aplicao-alvo}
Para realizar o processo de decomposição da aplicação monolítica foram utilizados os princípios do \textit{Domain-Driven Design}~\cite{evans2009domain}, desta forma serão definidos os domínios juntamente com os seus delimitadores.
A aplicação alvo possui um alto nível de criticidade, desta forma será preciso realizar uma migração gradativa na utilização dos microsserviços decompostos. Haverá um chaveamento por tipo de terminal de atendimento e números de contas, desta forma será possível ter um controle maior sobre os possíveis clientes impactados. Os microsserviços resultantes da decomposição foram construídos utilizando a linguagem \textit{Java} e o \textit{framework Spring}. A escolha das tecnologias utilizadas se deu através do conhecimento e experiência comum entre os membros da equipe técnica. A comunicação entre os microsserviços é feita através do protocolo \textit{HTTP}, seguindo o padrão \textit{REST}.

Para a decomposição da base de dados, foi utilizado o \textit{SQL Server}, onde a estrutura de dados foi implementada para ser migrada parcialmente, tendo em vista a grande dependência dos demais sistemas com as tabelas da aplicação alvo.

\subsubsection{Testes de carga}
Os testes de cargas serão aplicados utilizando a ferramenta \textit{Gatling}, a qual é uma ferramenta de alta performance, uso simples e fácil manutenção. A ferramenta \textit{Gatling} suporta protocolo HTTP, a qual será a forma de comunicação presente nas duas arquiteturas da aplicação. Os cenários serão escritos utilizando a linguagem \textit{Scala}, o que torna necessário conhecer o básico da linguagem. Contudo, ao final dos testes é gerado um relatório contendo informações como: usuários ativos durante a simulação, requisições por segundo, respostas por segundo e uma variedade de percentuais sobre os tempos de resposta das requisições. A execução do teste de carga simulou 4 quantidades diferentes de requisições por segundo durante determinado período. A~\autoref{table:teste-carga} lista as simulações executadas para o teste de carga.

\begin{table}[h]
\centering
\caption{Tabela simulações do teste de carga}
\label{table:teste-carga}
\begin{tabular}{|c|l|} 
\hline
\begin{tabular}[c]{@{}c@{}}\textbf{Quantidade requisições}\\\textbf{por segundo}\end{tabular} & \textbf{Duração}  \\ 
\hline
10                                                                                            & 10 minutos        \\ 
\hline
20                                                                                            & 10 minutos        \\ 
\hline
30                                                                                            & 10 minutos        \\ 
\hline
40                                                                                            & 10 minutos        \\
\hline
\end{tabular}
\end{table}

\subsection{Variáveis e Método de Quantificação}
\label{sec:variaveis-metodo-quantificacao}

\textit{Variável dependente na primeira hipótese.} A variável dependente na primeira hipótese será a de métricas de modularidade de software descritas na \autoref{table:metricas-modularidade-software}, onde o código será avaliado sob diversos aspectos da modularidade de software. O cálculo dessa variável permite estudar o impacto variável de cada métrica.

\textit{Variável dependente na segunda hipótese.} A variável dependente na segunda hipótese será a de métricas de performance descritas na \autoref{table:metricas-performance}. O cálculo dessa variável permite estudar o impacto no uso de recursos de hardware em ambas abordagens arquiteturais, de acordo com uma carga de uso definida.

\textit{Variável independente.} A variável independente dos hipóteses 1 e 2 será o estilo arquitetural monolítico e o baseado em microsserviços.

\subsection{Métricas Selecionadas}
\label{sec:metricas-selecionadas}
Neste estudo serão utilizados os conjuntos de métricas \textit{Acoplamento} e \textit{Coesão} para avaliar a modularidade de software. Para avaliação de performance serão utilizadas métricas \textit{Consumo de CPU} e \textit{Consumo de memória}.

A~\autoref{table:metricas-modularidade-software} apresenta as métricas selecionadas para quantificar variáveis de modularidade, sendo acoplamento e coesão. A seleção das métricas foi feita com base em estudos empíricos anteriores~\cite{garcia2006modularizing, juliano2014detection} que comprovam a validade destas métricas para análise de modularidade de software. As métricas utilizadas foram:

\begin{itemize}
    \item \textbf{CBO:} Esta métrica mede quantas classes uma determinada classe
depende. Em Orientação a Objetos, o baixo acoplamento entre objetos indica um bom grau de modularidade~\cite{chidamber1994metrics}.
    \item \textbf{DIT:} Esta métrica é definida como o comprimento máximo do nó até a raiz da árvore. Quanto mais profunda uma classe está na hierarquia, maior o número de métodos que ela provavelmente herdará, tornando mais complexo prever seu comportamento~\cite{chidamber1994metrics}.
    \item \textbf{WMC:} Métrica que mede a complexidade da classe, obtida em termos da complexidade de cada um de seus métodos. Valor maior indica maior tempo e esforço para desenvolver e manter a classe. Quanto maior o número de métodos de uma classe, maior será o impacto potencial sobre os filhos, pois estes herdarão os métodos da classe pai~\cite{chidamber1994metrics}.
    \item \textbf{NOC:} Métrica que mede a largura da hierarquia de uma classe. Quanto maior o número, maior o reaproveitamento, entretanto quanto maior o número de filhos, maior a probabilidade de abstração inadequada da classe pai, além da possibilidade de necessitar de mais testes dos métodos dessa classe~\cite{chidamber1994metrics}.
    \item \textbf{LCOM:} Métrica que mede a falta de coesão de uma classe, selecionando todos os pares de métodos de uma classe e verificando se esses compartilham algum atributo. Quanto maior o número, maior é a falta de coesão, aumentando assim a complexidade, levando a um provável aumento do número de defeitos injetados no software~\cite{chidamber1994metrics}.
\end{itemize}

\begin{table*}
\centering
\caption{Tabela métricas de modularidade  de software~\cite{garcia2006modularizing}}
\small
\label{table:metricas-modularidade-software}
\begin{tabular}{|l|l|l|} 
\hline
\textbf{Atributos}           & \textbf{\textbf{Métricas}}                                                          & \textbf{Definições}                                                                                                                                                                                    \\ 
\hline
\multirow{4}{*}{Acoplamento} & Coupling betweenobjects (CBO)                                                       & \begin{tabular}[c]{@{}l@{}}Esta métrica mede quantas classes uma\\determinada classe depende.\end{tabular}                                                                                             \\ 
\cline{2-3}
                             & Depth inheritancetree(DIT)                                                          & \begin{tabular}[c]{@{}l@{}}Esta métrica é definida como o\\comprimento máximo do nó até a raiz\\da árvore.\end{tabular}                                                                                \\ 
\cline{2-3}
                             & \begin{tabular}[c]{@{}l@{}}Weighted Methods \\per Class (WMC)\end{tabular}          & \begin{tabular}[c]{@{}l@{}}Métrica que mede a complexidade da\\classe, obtida em termos da complexidade\\de cada um de seus métodos.\end{tabular}                                                       \\ 
\cline{2-3}
                             & \begin{tabular}[c]{@{}l@{}}Number of Children\\(NOC)\end{tabular}                   & \begin{tabular}[c]{@{}l@{}}Métrica que mede a largura da hierarquia\\de uma classe.\end{tabular}                                                                                                       \\ 
\hline
Coesão                       & \begin{tabular}[c]{@{}l@{}}Lack of Cohesion \\in Methods \\(LCOM/LOCM)\end{tabular} & \begin{tabular}[c]{@{}l@{}}Esta métrica mede a falta de coesão de\\uma classe, selecionando todos os pares\\de métodos de uma classe e verificando\\se esses compartilham algum atributo.\end{tabular}  \\
\hline
\end{tabular}
\end{table*}

Para comparação de performance entre a arquitetura monolítica e a arquitetura baseada em microsserviços, foram selecionadas as métricas de consumo de CPU e consumo de memória, visando determinar o desempenho de cada arquitetura~\cite{artigo4}. O consumo de CPU se refere à porcentagem de unidade do recurso CPU virtual que está sendo usado pelo \textit{Kubernetes}~\cite{kubernetes_cpu_2021}. O consumo de memória se refere à porcentagem de memória que foi utilizada na execução do processo~\cite{kubernetes_memory_2021}.

\begin{table*}[H]
\centering
\caption{Tabela métricas de performance}
\label{table:metricas-performance}
\begin{tabular}{|l|l|} 
\hline
\textbf{Métricas}           & \textbf{Definições}                                                                                                                                \\ 
\hline
\textbf{Consumo de CPU}     & \begin{tabular}[c]{@{}l@{}}Consumo do recurso de CPU durante os testes de carga. \\Quanto menor for o consumo de CPU, melhor.\end{tabular}         \\ 
\hline
\textbf{Consumo de memória} & \begin{tabular}[c]{@{}l@{}}Consumo do recurso de memória durante os testes de carga.\\Quanto menor for o consumo de memória, melhor.\end{tabular}  \\
\hline
\end{tabular}
\end{table*}

\subsection{Processo Experimental}
\label{sec:procedimento-experimental}

\begin{figure}[H]
  \centering
  \includegraphics[scale=0.5]{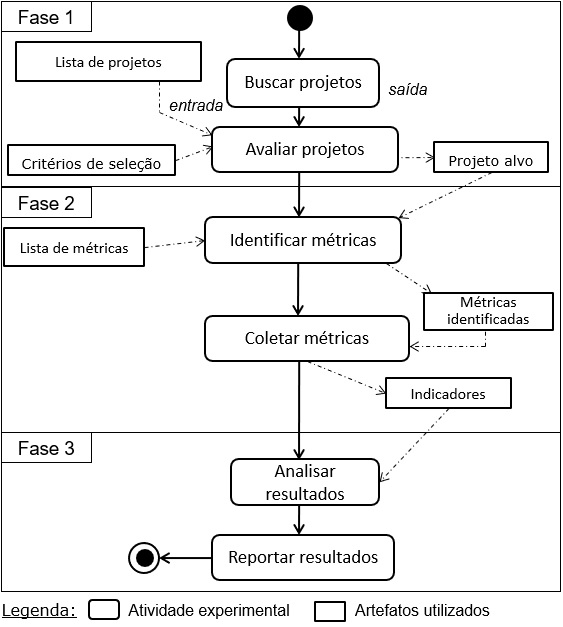}
  \caption{Processo experimental}
  \label{fig:processo-experimental}
\end{figure}

O processo experimental definido foi baseado em estudos empíricos publicados~\cite{farias2014effects, farias2015evaluating,d2020effects,farias2016empirical}, onde é identificado um conjunto de atividades, organizadas em três fases (Figura ~\ref{fig:processo-experimental}).

A Figura~\ref{fig:processo-experimental} mostra através de um processo experimental como as três fases foram organizadas. As atividades são descritas a seguir:
\begin{itemize}
    \item \textbf{Buscar projetos:} foi realizada a busca de projetos no repositório de códigos da \textit{Cooperativa Utile}.
    \item \textbf{Avaliar projetos:} foi realizada a avaliação dos projetos, para averiguar quais seriam os mais aptos a serem utilizados no estudo de caso. A seleção se baseou nas características definidas na Seção~\ref{sec:selecao-aplicacao-alvo}.
    \item \textbf{Mapear fluxo da aplicação:} identificado o fluxo da aplicação alvo e criação dos testes de carga.
    \item \textbf{Executar testes de carga:} testes de carga executados e os dados resultantes foram coletados.
    \item \textbf{Avaliar resultados:} avaliação dos resultados dos testes de carga usando o método de quantificação descrito na Seção~\ref{sec:variaveis-metodo-quantificacao}.
\end{itemize}

\subsection{Procedimento de Análise}
\label{sec:procedimento-analise}

A análise quantitativa dos dados de modularidade de software, será realizada através dos dados coletados pela ferramenta \textit{Understand}\footnote{Understand: https://www.scitools.com/}. Onde para a análise de distribuição dos dados de cada métrica, serão utilizados os seguintes métodos estatísticos: desvio padrão, valor máximo, mediana, média e percentual de variação entre as médias. Outros estudos usam desta abordagem para analisar métricas de software~\cite{juliano2014detection}.

A análise quantitativa dos dados refentes à análise de performance, será realizada através da estatística descritiva para analisar sua distribuição normal~\cite{wohlin2012experimentation} e inferência estatística para testar as hipóteses. O nível de significância dos testes de hipótese serão $\alpha$ = 0,05. Para testar as hipóteses, será aplicado o teste T das amostras~\cite{lilja2005measuring}.

\section{Resultados}
\label{ch:resultados}
Esta seção tem como objetivo apresentar os resultados referentes as questões de pesquisas formuladas na~\autoref{ch:metodologia}. A Seção~\ref{sec:analise-performance} discute os resultados obtidos referentes a performance de hardware. Na Seção~\ref{sec:analise-metricas-modularidade-software} discute-se os resultados obtidos referentes a modularidade de software. Por fim a Seção~\ref{sec:discussao-resultados} apresenta uma discussão adicional sobre os resultados obtidos.

\subsection{Análise de performance}
\label{sec:analise-performance}
A aplicação alvo da empresa hipotética~\textit{Cooperativa Utile}, do ramo financeiro, foi projetada para atender operações realizadas no terminal de autoatendimento. Considerando as aplicações monolíticas já existentes na empresa, o valor médio para consumo de CPU é de 2,5 vCPU, já o valor médio de consumo de memória é de 1536 MiB.

\subsubsection{Consumo de Memória}
\textbf{Estatística descritiva.} A~\autoref{table:resultados-consumo-memoria} apresenta os dados referentes a variável do consumo de memória. Embora esse valor baixo, é necessário verificar se esse valor obtido possui diferença estatisticamente relevante ao limite aceitável hipoteticamente definido pela \textit{Cooperativa Utile}. Para isso será testada a hipótese no parágrafo à frente.

\begin{table}[H]
\centering
\caption{Tabela resultados consumo de Memória (medidas em MebiByte - MiB)}
\label{table:resultados-consumo-memoria}
\small
\begin{tabular}{|l|l|l|l|l|} 
\hline
\begin{tabular}[c]{@{}l@{}}\textbf{Tamanho da}\\\textbf{amostra}\end{tabular} & \textbf{Menor} & \textbf{Maior} & \textbf{Média} & \textbf{Desvio padrão}  \\ 
\hline
\multicolumn{1}{|r|}{84}                                                      & 479,75         & 788,84         & 627,68         & 95,09                   \\
\hline
\end{tabular}
\end{table}

\textbf{Teste de hipótese questão de pesquisa 1}.

\begin{tcolorbox}[colback=blue!9]
\textbf{Resultados observados 1:} 
\textit{O consumo médio de memória (M = 627,68, SD = 95,09) foi menor do que o consumo médio padrão das aplicações monolíticas de 1536,00, uma diferença média estatisticamente significativa de 908,32, IC de 95\% [607,05, 648,32], t (84) = -87,54, p-value = 0,001.} %
\end{tcolorbox}

\subsubsection{Consumo de CPU}
\textbf{Estatística descritiva.} A~\autoref{table:resultados-consumo-cpu} apresenta os dados referentes à variável do consumo de CPU. Embora esse valor seja  baixo, é necessário verificar se esse valor obtido possui diferença estatisticamente relevante ao limite aceitável hipoteticamente definido pela \textit{Cooperativa Utile}. Para isso, será testada a hipótese no parágrafo à frente.
\begin{table}[H]
\centering
\caption{Tabela resultados consumo de CPU (medidas em vCPU)}
\label{table:resultados-consumo-cpu}
\small
\begin{tabular}{|l|l|l|l|l|} 
\hline
\begin{tabular}[c]{@{}l@{}}\textbf{Tamanho da}\\\textbf{amostra}\end{tabular} & \textbf{Menor} & \textbf{Maior} & \textbf{Média} & \textbf{Desvio padrão}  \\ 
\hline
\multicolumn{1}{|r|}{84}                                                      & 0,14           & 0,92           & 0,45~ ~~    & 0,16                 \\
\hline
\end{tabular}
\end{table}

A Tabela~\ref{table:resultados-comparacao-hipotetica-consumo-cpu} apresenta os resultados do teste de hipótese para análise do consumo de \textit{CPU}. Os valores apontam que o \textit{p} < 0,05 indicando que há uma diferença estatisticamente relevante, onde o uso da arquitetura baseada em microsserviços apresenta uma diferença significativa, entre o valor de limite aceitável e o valor apresentado. Portanto, a arquitetura baseada em microsserviços se configura como sendo uma arquitetura promissora, corroborando com outros estudos~\cite{artigo4,artigo5} que evidenciam menor consumo dos recursos de hardware.

\textbf{Teste de hipótese questão de pesquisa 1}.
\begin{tcolorbox}[colback=blue!9]
\label{table:resultados-comparacao-hipotetica-consumo-cpu}
\textbf{Resultados observados 2:} 
\textit{O percentual de consumo médio de CPU (M = 0,45, SD = 0,16) foi menor do que o percentual de consumo médio padrão dos microsserviços de 2,50, uma diferença média estatisticamente significativa de 2,05, IC de 95\% [0,42, 0,49], t (84) = -116,7, p-value = 0,001.} %
\end{tcolorbox}

\subsection{Análise das métricas de modularidade de software}
\label{sec:analise-metricas-modularidade-software}

A Tabela~\ref{table:resultados-metricas-modularidade} traz os indicadores estatísticos sobre o resultado das métricas coletadas, referentes à modularidade de software. Apresenta os resultados dos efeitos da arquitetura baseada em microsserviços em comparação à arquitetura monolítica, através dos atributos de Acoplamento e Coesão, incluindo o desvio padrão, mediana, média e percentual de variação entre as médias. Este conjunto de métricas visa identificar a modularidade do software, já a~\autoref{table:valores-referencia-metricas-ck} apresenta valores de referência para as métricas~\cite{chidamber1994metrics}, tendo como categorias: Baixo, Alto e Anomalia.


\begin{table*}[h]
\centering
\caption{Tabela resultados das métricas de modularidade de software}
\label{table:resultados-metricas-modularidade}
\scriptsize
\begin{tabular}{|l|l|l|r|c|r|r|c|} 
\hline
\textbf{Atributos}           & \textbf{\textbf{Métricas}} & \textbf{Arquitetura} & \multicolumn{1}{l|}{\textbf{Desvio Padrão}} & \textbf{Máximo} & \multicolumn{1}{l|}{\textbf{Mediana}} & \multicolumn{1}{l|}{\textbf{Média}} & \multicolumn{1}{l|}{\textbf{\% de variação}}  \\ 
\hline
\multirow{8}{*}{Acoplamento} & \multirow{2}{*}{CBO}       & Monolítica           & 12,53                                       & 121             & 5                                     & 10,06                               & \multirow{2}{*}{30,76\%}                      \\ 
\cline{3-7}
                             &                            & Microsserviços       & 7,22                                        & 38              & 6                                     & 7,70                                &                                               \\ 
\cline{2-8}
                             & \multirow{2}{*}{DIT}       & Monolítica           & 0,92                                        & 4               & 1                                     & 1,62                                & \multirow{2}{*}{76,67\%}                      \\ 
\cline{3-7}
                             &                            & Microsserviços       & 0,51                                        & 3               & 1                                     & 0,92                                &                                               \\ 
\cline{2-8}
                             & \multirow{2}{*}{WMC}       & Monolítica           & 82,81                                       & 2327            & 7                                     & 19,34                               & \multirow{2}{*}{209,85\%}                     \\ 
\cline{3-7}
                             &                            & Microsserviços       & 15,79                                       & 231             & 2,5                                   & 6,24                                &                                               \\ 
\cline{2-8}
                             & \multirow{2}{*}{NOC}       & Monolítica           & 4,60                                        & 86              & 0                                     & 0,51                                & \multirow{2}{*}{121,35\%~~}                   \\ 
\cline{3-7}
                             &                            & Microsserviços       & 0,62                                        & 4               & 0                                     & 0,23                                &                                               \\ 
\hline
\multirow{2}{*}{Coesão}      & \multirow{2}{*}{LCOM}      & Monolítica           & 37,84                                       & 100             & 61                                    & 48,49                               & \multirow{2}{*}{129,32\%}                     \\ 
\cline{3-7}
                             &                            & Microsserviços       & 30,02                                       & 100             & 0                                     & 21,15                               &                                               \\
\hline
\end{tabular}
\end{table*}

\begin{table}[H]
\centering
\caption{Valores referência para as métricas CK em softwares desenvolvidos na linguagem Java~\cite{juliano2014detection}}
\label{table:valores-referencia-metricas-ck}
\small
\begin{tabular}{|l|c|c|c|c|c|c|} 
\cline{2-7}
\multicolumn{1}{l|}{} & \multicolumn{1}{l|}{\textbf{DIT}} & \multicolumn{1}{l|}{\textbf{NOC}} & \multicolumn{1}{l|}{\textbf{CBO}} & \multicolumn{1}{l|}{\textbf{RFC}} & \multicolumn{1}{l|}{\textbf{LCOM}} & \multicolumn{1}{l|}{\textbf{WMC}}  \\ 
\hline
\textbf{Baixo}        & 0                                 & 0                                 & 0                                 & 0                                 & 0                                  & 230                                \\ 
\hline
\textbf{Alto}         & 2,5                               & 6,4                               & 13,6                              & 35,5                              & 92,7                               & 253,1                              \\ 
\hline
\textbf{Anomalia}     & 3                                 & 8                                 & 18                                & 46                                & 120                                & 329                                \\
\hline
\end{tabular}
\end{table}

A categoria de Acoplamento traz as seguintes métricas para comparação: \textit{CBO}, \textit{DIT}, \textit{WMC} e \textit{NOC}. A arquitetura baseada em microsserviços apresentou resultados melhores em comparação à arquitetura monolítica. Isso pode ser identificado através da diferença entre as médias de cada arquitetura, representada pelo percentual de variação, sendo 30,76\%, 76,67\%, 209,85\% e 121,35\% respectivamente.

\textbf{CBO.} O percentual de variação de 30,76\% para a arquitetura de microsserviços, indica um melhor grau de modularidade do código, onde os princípios da Orientação a Objetos foram melhores aplicados. Ambas arquiteturas obtiveram médias classificadas entre Baixo e Alto, segundo os valores de referência da~\autoref{table:valores-referencia-metricas-ck}. Entretanto, ambas arquiteturas tiverem classes com valores categorizados como Anomalia.

\textbf{DIT.} A arquitetura de microsserviços tendo um percentual de variação menor, indica menos complexidade e também a possibilidade de menos reutilização de código por meio de herança. Ambas arquiteturas obtiveram médias classificadas entre Baixo e Alto, segundo os valores de referência da~\autoref{table:valores-referencia-metricas-ck}. Contudo, os resultados obtidos indicam que ambas arquiteturas tiveram classes com valores acima do categorizado como Anomalia, indicando que ambos projetos possuem classes problemáticas.

\textbf{WMC.} A arquitetura monolítica teve percentual de variação superior em 209,85\%, indicando classes com maior número de métodos e complexidades entre eles. Além do percentual de variação em comparação à arquitetura de microsserviços, esta métrica obteve classes com valores superiores ao categorizado como Anomalia.

\textbf{NOC.} A arquitetura monolítica apresentou um número maior desta métrica, a qual obteve classes com valor máximo de 86. Um alto valor desta métrica indica que há probabilidade da classe apresentar defeitos, devido à grande hierarquia das classes~\cite{juliano2014detection}. 

\textbf{LCOM.} Os resultados coletados desta métrica trazem uma redução de 129,32\% para a arquitetura de microsserviços em comparação com a arquitetura monolítica. Entretanto, ambas arquiteturas obtiveram valores máximos de 100, podendo ser classificadas com categoria Alto, segundo a~\autoref{table:valores-referencia-metricas-ck} com valores de referência.

Estes resultados eram esperados, pois a aplicação monolítica apresentou um número elevado de classes e linhas de códigos. Isso pode ser explicado pelo fato da aplicação monolítica possuir um conjunto de funcionalidades maiores, consequentemente necessitando de um número maior de classes e linhas de código, outros estudos indicam estes resultados~\cite{policyEnforcementUponSoftware}. Entretanto, a aplicação baseada em microsserviços foi resultado de um processo de decomposição da aplicação monolítica, onde foram extraídas as funcionalidades necessárias para o produto, isso facilitou um melhor entendimento do domínio deste produto~\cite{evans2009domain}, onde foram melhor aplicados os conceitos da linguagem orientada a objetos, diminuindo a complexidade do código. Esta última afirmação se fundamenta principalmente através dos melhores resultados para as métricas WMC e CBO, as quais foram utilizadas com maior chance de acerto em outros estudos~\cite{juliano2014detection} para prever classes com propensão a erros.

\begin{tcolorbox}[colback=blue!9]
\textbf{Resultados observados 3:} 
\textit{As métricas de acoplamento e coesão obtiveram valores médios considerados baixos para ambas arquiteturas~\cite{juliano2014detection}. No entanto, a aplicação com arquitetura baseada em microsserviços, obteve valores menores em todas as métricas, evidenciando um menor acoplamento e maior coesão.} %
\end{tcolorbox}

\subsection{Discussão}
\label{sec:discussao-resultados}
\textbf{Modularidade de software.} A arquitetura monolítica obteve bons resultados considerando os valores de referência. Entretanto, ao analisar os resultados obtidos através dos conjuntos de métricas definidas, identificou-se que a arquitetura baseada em microsserviços obteve melhores resultados de forma geral. O baixo acoplamento indica um aumento na qualidade do \textit{software} e possivelmente maior reuso dos componentes~\cite{kazman1996scenario}.

\textbf{Análise de performance.} 
A arquitetura monolítica obteve valores elevados em comparação com a arquitetura baseada em microsserviços. Os valores de consumo de memória para os microsserviços indicam uma redução considerável, tendo em vista o contexto de software reduzido e uma menor dependência de outros softwares. O consumo de CPU e memória, inferiores na aplicação com arquitetura baseada em microsserviços, indicam menor carga de processamento computacional.

\subsection{Limitações do Estudo}
\label{sec:limitacoes}
O estudo de caso reportado se trata de um estudo inicial que explora um assunto pouco investigado na literatura. Desta forma, o estudo possui algumas limitações que devem ser consideradas. A aplicação monolítica considerada, possui diversas funcionalidades implementadas, além da funcionalidade alvo do estudo, podendo ter distorcido os resultados coletados. Tal consideração corrobora com o fato de ser uma aplicação antiga, onde não adotaram boas práticas e todos os conceitos da orientação a objetos. Sendo esse um dos motivos pelo qual houve um grande percentual de variação entre as métricas de modularidade de software. Outra dificuldade foi encontrada para coletar as métricas de consumo de memória e consumo de CPU, pela aplicação monolítica, onde não foi possível coletar devido a problemas técnicos no servidor onde os testes de carga foram executados. Neste caso, foram considerados valores baseados na experiência da equipe de desenvolvimento.

\section{Conclusão e Trabalhos Futuros}
\label{ch:conclusao-trabalhos-futuros}

A arquitetura de microsserviços vem sendo adotada na indústria como forma de modernização de aplicações legadas. Surge como alternativa para a arquitetura monolítica pois, traz maior escalabilidade e manutenibilidade da aplicação. No entanto, não há muitos estudos evidenciando os impactos da adoção deste estilo arquitetural em relação ao monolítico. Neste sentido, o estudo atual reportou um estudo de caso inicial com o propósito de comparar a arquitetura monolítica e a arquitetura de microsserviços. O estudo reportado procurou avaliar os impactos da utilização de ambas arquiteturas, através de métricas computacionais como, consumo de CPU, consumo de memória, além de métricas para medir a modularidade do software.

No estudo atual, as métricas selecionadas são avaliadas em uma aplicação financeira real, que atende as operações de saque, dos terminais de autoatendimento, da empresa fictícia \textit{Cooperativa Utile}.

As descobertas indicam que o uso da arquitetura de microsserviços, apresentou bons resultados quanto as métricas de acoplamento e coesão. Contudo, ficou evidente a necessidade de um conjunto maior de métricas, para avaliar outros aspectos das aplicações, como por exemplo o acoplamento entre microsserviços.

As métricas computacionais para avaliar a performance tiveram valores hipotetizados para as métricas de consumo de CPU e consumo de memória. Novamente a arquitetura de microsserviços teve bons resultados em comparação a arquitetura monolítica. No entanto, mostrou-se necessário novas métricas para melhor avaliar a performance, como por exemplo a latência e a taxa de transferência~\cite{artigo5}. 

A realização deste estudo, trouxe contribuições científicas referentes ao tema de decomposição de aplicações monolíticas em microsserviços. O uso de uma aplicação real como estudo de caso, aumentou o conhecimento empírico gerado sobre o tema.

Como trabalhos futuros, pretende-se: (1) definir um conjunto maior de métricas, visando aumentar a perspectiva de análise da modularidade do software e performance; (2) coletar mais dados em ambas arquiteturas, para realizar uma analise estatística sem a necessidade de hipotetizar os valores. Este trabalho pode ser tido como sendo um estudo inicial, de uma sequência de estudos de caso mais robustos, relacionados aos impactos da decomposição de aplicações monolíticas para microsserviços.

\bibliographystyle{ACM-Reference-Format}
\bibliography{sample-sigconf}


\begin{thebibliography}{48}


\ifx \showCODEN    \undefined \def \showCODEN     #1{\unskip}     \fi
\ifx \showDOI      \undefined \def \showDOI       #1{#1}\fi
\ifx \showISBNx    \undefined \def \showISBNx     #1{\unskip}     \fi
\ifx \showISBNxiii \undefined \def \showISBNxiii  #1{\unskip}     \fi
\ifx \showISSN     \undefined \def \showISSN      #1{\unskip}     \fi
\ifx \showLCCN     \undefined \def \showLCCN      #1{\unskip}     \fi
\ifx \shownote     \undefined \def \shownote      #1{#1}          \fi
\ifx \showarticletitle \undefined \def \showarticletitle #1{#1}   \fi
\ifx \showURL      \undefined \def \showURL       {\relax}        \fi
\providecommand\bibfield[2]{#2}
\providecommand\bibinfo[2]{#2}
\providecommand\natexlab[1]{#1}
\providecommand\showeprint[2][]{arXiv:#2}

\bibitem[kub(2021a)]%
        {kubernetes_cpu_2021}
 \bibinfo{year}{2021}\natexlab{a}.
\newblock \bibinfo{title}{Managing Resources for Containers}.
\newblock
\newblock
\urldef\tempurl%
\url{https://kubernetes.io/docs/concepts/configuration/manage-resources-containers/#meaning-of-cpu}
\showURL{%
\tempurl}


\bibitem[kub(2021b)]%
        {kubernetes_memory_2021}
 \bibinfo{year}{2021}\natexlab{b}.
\newblock \bibinfo{title}{Managing Resources for Containers}.
\newblock
\newblock
\urldef\tempurl%
\url{https://kubernetes.io/docs/concepts/configuration/manage-resources-containers/#meaning-of-memory}
\showURL{%
\tempurl}


\bibitem[Allman(2012)]%
        {allman2012managing}
\bibfield{author}{\bibinfo{person}{Eric Allman}.}
  \bibinfo{year}{2012}\natexlab{}.
\newblock \showarticletitle{Managing technical debt}.
\newblock \bibinfo{journal}{\emph{Commun. ACM}} \bibinfo{volume}{55},
  \bibinfo{number}{5} (\bibinfo{year}{2012}), \bibinfo{pages}{50--55}.
\newblock


\bibitem[Arun(2015)]%
        {arun}
\bibfield{author}{\bibinfo{person}{Arun}.} \bibinfo{year}{2015}\natexlab{}.
\newblock \showarticletitle{A First Look at Microservices}.
\newblock \bibinfo{journal}{\emph{Java Magazine}}  \bibinfo{volume}{sep/oct}
  (\bibinfo{year}{2015}).
\newblock


\bibitem[Asik and Selcuk(2017)]%
        {policyEnforcementUponSoftware}
\bibfield{author}{\bibinfo{person}{Tugrul Asik} {and}
  \bibinfo{person}{Yunus~Emre Selcuk}.} \bibinfo{year}{2017}\natexlab{}.
\newblock \showarticletitle{Policy enforcement upon software based on
  microservice architecture}. In \bibinfo{booktitle}{\emph{2017 IEEE 15th
  International Conference on Software Engineering Research, Management and
  Applications (SERA)}}. \bibinfo{pages}{283--287}.
\newblock
\urldef\tempurl%
\url{https://doi.org/10.1109/SERA.2017.7965739}
\showDOI{\tempurl}


\bibitem[Avgeriou et~al\mbox{.}(2016)]%
        {avgeriou2016managing}
\bibfield{author}{\bibinfo{person}{Paris Avgeriou}, \bibinfo{person}{Philippe
  Kruchten}, \bibinfo{person}{Ipek Ozkaya}, {and} \bibinfo{person}{Carolyn
  Seaman}.} \bibinfo{year}{2016}\natexlab{}.
\newblock \showarticletitle{Managing technical debt in software engineering
  (dagstuhl seminar 16162)}. In \bibinfo{booktitle}{\emph{Dagstuhl Reports}},
  Vol.~\bibinfo{volume}{6}. Schloss Dagstuhl-Leibniz-Zentrum fuer Informatik.
\newblock


\bibitem[Baumann et~al\mbox{.}(2009)]%
        {Baumann09}
\bibfield{author}{\bibinfo{person}{Andrew Baumann}, \bibinfo{person}{Paul
  Barham}, \bibinfo{person}{Pierre-Evariste Dagand}, \bibinfo{person}{Tim
  Harris}, \bibinfo{person}{Rebecca Isaacs}, \bibinfo{person}{Simon Peter},
  \bibinfo{person}{Timothy Roscoe}, \bibinfo{person}{Adrian Sch{\"u}pbach},
  {and} \bibinfo{person}{Akhilesh Singhania}.} \bibinfo{year}{2009}\natexlab{}.
\newblock \showarticletitle{The {M}ultikernel: a New {OS} Architecture for
  Scalable Multicore Systems}. In \bibinfo{booktitle}{\emph{Proc.\ of the 22nd
  ACM SIGOPS Symposium on Operating Systems Principles}}. \bibinfo{address}{Big
  Sky, Montana, USA}, \bibinfo{pages}{29--44}.
\newblock
\showISBNx{978-1-60558-752-3}
\urldef\tempurl%
\url{https://doi.org/10.1145/1629575.1629579}
\showDOI{\tempurl}


\bibitem[Bjørndal et~al\mbox{.}(2020)]%
        {artigo5}
\bibfield{author}{\bibinfo{person}{N Bjørndal}, \bibinfo{person}{Antonio
  Bucchiarone}, \bibinfo{person}{Manuel Mazzara}, \bibinfo{person}{Nicola
  Dragoni}, {and} \bibinfo{person}{Schahram Dustdar}.}
  \bibinfo{year}{2020}\natexlab{}.
\newblock \showarticletitle{Migration from Monolith to Microservices :
  Benchmarking a Case Study}.
\newblock  (\bibinfo{date}{03} \bibinfo{year}{2020}).
\newblock
\urldef\tempurl%
\url{https://doi.org/10.13140/RG.2.2.27715.14883}
\showDOI{\tempurl}


\bibitem[Brown et~al\mbox{.}(2010)]%
        {brown2010managing}
\bibfield{author}{\bibinfo{person}{Nanette Brown}, \bibinfo{person}{Yuanfang
  Cai}, \bibinfo{person}{Yuepu Guo}, \bibinfo{person}{Rick Kazman},
  \bibinfo{person}{Miryung Kim}, \bibinfo{person}{Philippe Kruchten},
  \bibinfo{person}{Erin Lim}, \bibinfo{person}{Alan MacCormack},
  \bibinfo{person}{Robert Nord}, \bibinfo{person}{Ipek Ozkaya},
  {et~al\mbox{.}}} \bibinfo{year}{2010}\natexlab{}.
\newblock \showarticletitle{Managing technical debt in software-reliant
  systems}. In \bibinfo{booktitle}{\emph{Proceedings of the FSE/SDP workshop on
  Future of software engineering research}}. \bibinfo{pages}{47--52}.
\newblock


\bibitem[Camargo et~al\mbox{.}(2016)]%
        {camargo2016abordagem}
\bibfield{author}{\bibinfo{person}{Andr{\'e} Stangarlin~de Camargo}
  {et~al\mbox{.}}} \bibinfo{year}{2016}\natexlab{}.
\newblock \showarticletitle{Uma abordagem para testes de desempenho de
  microservices}.
\newblock  (\bibinfo{year}{2016}).
\newblock


\bibitem[Carbonera et~al\mbox{.}(2020)]%
        {carbonera2020software}
\bibfield{author}{\bibinfo{person}{Carlos~Eduardo Carbonera},
  \bibinfo{person}{Kleinner Farias}, {and} \bibinfo{person}{Vinicius
  Bischoff}.} \bibinfo{year}{2020}\natexlab{}.
\newblock \showarticletitle{Software development effort estimation: a
  systematic mapping study}.
\newblock \bibinfo{journal}{\emph{IET Software}} \bibinfo{volume}{14},
  \bibinfo{number}{4} (\bibinfo{year}{2020}), \bibinfo{pages}{328--344}.
\newblock


\bibitem[Chidamber and Kemerer(1994)]%
        {chidamber1994metrics}
\bibfield{author}{\bibinfo{person}{Shyam~R Chidamber} {and}
  \bibinfo{person}{Chris~F Kemerer}.} \bibinfo{year}{1994}\natexlab{}.
\newblock \showarticletitle{A metrics suite for object oriented design}.
\newblock \bibinfo{journal}{\emph{IEEE Transactions on software engineering}}
  \bibinfo{volume}{20}, \bibinfo{number}{6} (\bibinfo{year}{1994}),
  \bibinfo{pages}{476--493}.
\newblock


\bibitem[Desai(2016)]%
        {desai2016survey}
\bibfield{author}{\bibinfo{person}{Prashant~Ramchandra Desai}.}
  \bibinfo{year}{2016}\natexlab{}.
\newblock \showarticletitle{A survey of performance comparison between virtual
  machines and containers}.
\newblock \bibinfo{journal}{\emph{Int. J. Comput. Sci. Eng}}
  \bibinfo{volume}{4}, \bibinfo{number}{7} (\bibinfo{year}{2016}),
  \bibinfo{pages}{55--59}.
\newblock


\bibitem[Dragoni et~al\mbox{.}(2017)]%
        {dragoni2017microservices}
\bibfield{author}{\bibinfo{person}{Nicola Dragoni}, \bibinfo{person}{Saverio
  Giallorenzo}, \bibinfo{person}{Alberto~Lluch Lafuente},
  \bibinfo{person}{Manuel Mazzara}, \bibinfo{person}{Fabrizio Montesi},
  \bibinfo{person}{Ruslan Mustafin}, {and} \bibinfo{person}{Larisa Safina}.}
  \bibinfo{year}{2017}\natexlab{}.
\newblock \showarticletitle{Microservices: yesterday, today, and tomorrow}.
\newblock \bibinfo{journal}{\emph{Present and ulterior software engineering}}
  (\bibinfo{year}{2017}), \bibinfo{pages}{195--216}.
\newblock


\bibitem[D’Avila et~al\mbox{.}(2020)]%
        {d2020effects}
\bibfield{author}{\bibinfo{person}{Leandro~Ferreira D’Avila},
  \bibinfo{person}{Kleinner Farias}, {and} \bibinfo{person}{Jorge
  Luis~Vict{\'o}ria Barbosa}.} \bibinfo{year}{2020}\natexlab{}.
\newblock \showarticletitle{Effects of contextual information on maintenance
  effort: a controlled experiment}.
\newblock \bibinfo{journal}{\emph{Journal of Systems and Software}}
  \bibinfo{volume}{159} (\bibinfo{year}{2020}), \bibinfo{pages}{110443}.
\newblock


\bibitem[Evans(2009)]%
        {evans2009domain}
\bibfield{author}{\bibinfo{person}{Eric Evans}.}
  \bibinfo{year}{2009}\natexlab{}.
\newblock \bibinfo{booktitle}{\emph{Domain-driven design: atacando as
  complexidades no cora{\c{c}}{\~a}o do software}}.
\newblock \bibinfo{publisher}{Alta Books}.
\newblock


\bibitem[Farias(2016)]%
        {farias2016empirical}
\bibfield{author}{\bibinfo{person}{Kleinner Farias}.}
  \bibinfo{year}{2016}\natexlab{}.
\newblock \showarticletitle{Empirical evaluation of effort on composing design
  models}.
\newblock \bibinfo{journal}{\emph{arXiv preprint arXiv:1610.09012}}
  (\bibinfo{year}{2016}).
\newblock


\bibitem[Farias et~al\mbox{.}(2014)]%
        {farias2014effects}
\bibfield{author}{\bibinfo{person}{Kleinner Farias},
  \bibinfo{person}{Alessandro Garcia}, {and} \bibinfo{person}{Carlos Lucena}.}
  \bibinfo{year}{2014}\natexlab{}.
\newblock \showarticletitle{Effects of stability on model composition effort:
  an exploratory study}.
\newblock \bibinfo{journal}{\emph{Software \& Systems Modeling}}
  \bibinfo{volume}{13}, \bibinfo{number}{4} (\bibinfo{year}{2014}),
  \bibinfo{pages}{1473--1494}.
\newblock


\bibitem[Farias et~al\mbox{.}(2015)]%
        {farias2015evaluating}
\bibfield{author}{\bibinfo{person}{Kleinner Farias},
  \bibinfo{person}{Alessandro Garcia}, \bibinfo{person}{Jon Whittle},
  \bibinfo{person}{Christina von Flach~Garcia Chavez}, {and}
  \bibinfo{person}{Carlos Lucena}.} \bibinfo{year}{2015}\natexlab{}.
\newblock \showarticletitle{Evaluating the effort of composing design models: a
  controlled experiment}.
\newblock \bibinfo{journal}{\emph{Software \& Systems Modeling}}
  \bibinfo{volume}{14}, \bibinfo{number}{4} (\bibinfo{year}{2015}),
  \bibinfo{pages}{1349--1365}.
\newblock


\bibitem[Farias et~al\mbox{.}(2018)]%
        {farias2018uml}
\bibfield{author}{\bibinfo{person}{Kleinner Farias}, \bibinfo{person}{Lucian
  Gon{\c{c}}ales}, \bibinfo{person}{Vinicius Bischoff},
  \bibinfo{person}{Bruno~Carreiro da Silva}, \bibinfo{person}{Everton~T
  Guimar{\~a}es}, {and} \bibinfo{person}{Jacob Nogle}.}
  \bibinfo{year}{2018}\natexlab{}.
\newblock \showarticletitle{On the UML use in the Brazilian industry: A state
  of the practice survey (S).}. In \bibinfo{booktitle}{\emph{SEKE}}.
  \bibinfo{pages}{372--371}.
\newblock


\bibitem[Fowler(2015)]%
        {fowler_monolith_first}
\bibfield{author}{\bibinfo{person}{Martin Fowler}.}
  \bibinfo{year}{2015}\natexlab{}.
\newblock \bibinfo{title}{Monolith First}.
\newblock
\newblock
\urldef\tempurl%
\url{https://martinfowler.com/bliki/MonolithFirst.html}
\showURL{%
\tempurl}


\bibitem[Fowler et~al\mbox{.}(1997)]%
        {fowler1997refactoring}
\bibfield{author}{\bibinfo{person}{Martin Fowler}, \bibinfo{person}{Kent Beck},
  {and} \bibinfo{person}{W~Roberts Opdyke}.} \bibinfo{year}{1997}\natexlab{}.
\newblock \showarticletitle{Refactoring: Improving the design of existing
  code}. In \bibinfo{booktitle}{\emph{11th European Conference.
  Jyv{\"a}skyl{\"a}, Finland}}.
\newblock


\bibitem[Fowler and Lewis(2014)]%
        {fowler_microservices}
\bibfield{author}{\bibinfo{person}{Martin Fowler} {and} \bibinfo{person}{James
  Lewis}.} \bibinfo{year}{2014}\natexlab{}.
\newblock \bibinfo{title}{Microservices, a definition of this new architectural
  term}.
\newblock
\newblock
\urldef\tempurl%
\url{https://martinfowler.com/articles/microservices.html}
\showURL{%
\tempurl}


\bibitem[Garcia et~al\mbox{.}(2006)]%
        {garcia2006modularizing}
\bibfield{author}{\bibinfo{person}{Alessandro Garcia},
  \bibinfo{person}{Cl{\'a}udio Sant’Anna}, \bibinfo{person}{Eduardo
  Figueiredo}, \bibinfo{person}{Uir{\'a} Kulesza}, \bibinfo{person}{Carlos
  Lucena}, {and} \bibinfo{person}{Arndt von Staa}.}
  \bibinfo{year}{2006}\natexlab{}.
\newblock \showarticletitle{Modularizing design patterns with aspects: a
  quantitative study}.
\newblock In \bibinfo{booktitle}{\emph{Transactions on Aspect-Oriented Software
  Development I}}. \bibinfo{publisher}{Springer}, \bibinfo{pages}{36--74}.
\newblock


\bibitem[Gon{\c{c}}ales et~al\mbox{.}(2021)]%
        {gonccales2021measuring}
\bibfield{author}{\bibinfo{person}{Lucian Gon{\c{c}}ales},
  \bibinfo{person}{Kleinner Farias}, {and} \bibinfo{person}{Bruno~C da Silva}.}
  \bibinfo{year}{2021}\natexlab{}.
\newblock \showarticletitle{Measuring the cognitive load of software
  developers: An extended Systematic Mapping Study}.
\newblock \bibinfo{journal}{\emph{Information and Software Technology}}
  (\bibinfo{year}{2021}), \bibinfo{pages}{106563}.
\newblock


\bibitem[Gos and Zabierowski(2020)]%
        {artigo1}
\bibfield{author}{\bibinfo{person}{Konrad Gos} {and} \bibinfo{person}{Wojciech
  Zabierowski}.} \bibinfo{year}{2020}\natexlab{}.
\newblock \showarticletitle{The Comparison of Microservice and Monolithic
  Architecture}. In \bibinfo{booktitle}{\emph{2020 IEEE XVIth International
  Conference on the Perspective Technologies and Methods in MEMS Design
  (MEMSTECH)}}. IEEE, \bibinfo{pages}{150--153}.
\newblock


\bibitem[Gysel et~al\mbox{.}(2016)]%
        {gysel2016service}
\bibfield{author}{\bibinfo{person}{Michael Gysel}, \bibinfo{person}{Lukas
  K{\"o}lbener}, \bibinfo{person}{Wolfgang Giersche}, {and}
  \bibinfo{person}{Olaf Zimmermann}.} \bibinfo{year}{2016}\natexlab{}.
\newblock \showarticletitle{Service cutter: A systematic approach to service
  decomposition}. In \bibinfo{booktitle}{\emph{European Conference on
  Service-Oriented and Cloud Computing}}. Springer, \bibinfo{pages}{185--200}.
\newblock


\bibitem[Juliano et~al\mbox{.}(2014)]%
        {juliano2014detection}
\bibfield{author}{\bibinfo{person}{Renato~Correa Juliano},
  \bibinfo{person}{Bruno~AN Traven{\c{c}}olo}, {and} \bibinfo{person}{Michel~S
  Soares}.} \bibinfo{year}{2014}\natexlab{}.
\newblock \showarticletitle{Detection of Software Anomalies Using
  Object-oriented Metrics}. In \bibinfo{booktitle}{\emph{ICEIS (2)}}.
  \bibinfo{pages}{241--248}.
\newblock


\bibitem[J{\'u}nior et~al\mbox{.}(2021)]%
        {junior2021survey}
\bibfield{author}{\bibinfo{person}{Ed J{\'u}nior}, \bibinfo{person}{Kleinner
  Farias}, {and} \bibinfo{person}{Bruno Silva}.}
  \bibinfo{year}{2021}\natexlab{}.
\newblock \showarticletitle{A Survey on the Use of UML in the Brazilian
  Industry}. In \bibinfo{booktitle}{\emph{Brazilian Symposium on Software
  Engineering}}. \bibinfo{pages}{275--284}.
\newblock


\bibitem[Kazman et~al\mbox{.}(1996)]%
        {kazman1996scenario}
\bibfield{author}{\bibinfo{person}{Rick Kazman}, \bibinfo{person}{Gregory
  Abowd}, \bibinfo{person}{Len Bass}, {and} \bibinfo{person}{Paul Clements}.}
  \bibinfo{year}{1996}\natexlab{}.
\newblock \showarticletitle{Scenario-based analysis of software architecture}.
\newblock \bibinfo{journal}{\emph{IEEE software}} \bibinfo{volume}{13},
  \bibinfo{number}{6} (\bibinfo{year}{1996}), \bibinfo{pages}{47--55}.
\newblock


\bibitem[Knoche and Hasselbring(2018)]%
        {knoche2018using}
\bibfield{author}{\bibinfo{person}{Holger Knoche} {and}
  \bibinfo{person}{Wilhelm Hasselbring}.} \bibinfo{year}{2018}\natexlab{}.
\newblock \showarticletitle{Using microservices for legacy software
  modernization}.
\newblock \bibinfo{journal}{\emph{IEEE Software}} \bibinfo{volume}{35},
  \bibinfo{number}{3} (\bibinfo{year}{2018}), \bibinfo{pages}{44--49}.
\newblock


\bibitem[Lilja(2005)]%
        {lilja2005measuring}
\bibfield{author}{\bibinfo{person}{David~J Lilja}.}
  \bibinfo{year}{2005}\natexlab{}.
\newblock \bibinfo{booktitle}{\emph{Measuring computer performance: a
  practitioner's guide}}.
\newblock \bibinfo{publisher}{Cambridge university press}.
\newblock


\bibitem[Newman(2015)]%
        {newman2015building}
\bibfield{author}{\bibinfo{person}{Sam Newman}.}
  \bibinfo{year}{2015}\natexlab{}.
\newblock \bibinfo{booktitle}{\emph{Building microservices: designing
  fine-grained systems}}.
\newblock \bibinfo{publisher}{" O'Reilly Media, Inc."}.
\newblock


\bibitem[Oliveira et~al\mbox{.}(2018)]%
        {oliveira2018brcode}
\bibfield{author}{\bibinfo{person}{Anderson Oliveira},
  \bibinfo{person}{Vinicius Bischoff}, \bibinfo{person}{Lucian~Jos{\'e}
  Gon{\c{c}}ales}, \bibinfo{person}{Kleinner Farias}, {and}
  \bibinfo{person}{Matheus Segalotto}.} \bibinfo{year}{2018}\natexlab{}.
\newblock \showarticletitle{BRCode: An interpretive model-driven engineering
  approach for enterprise applications}.
\newblock \bibinfo{journal}{\emph{Computers in Industry}}  \bibinfo{volume}{96}
  (\bibinfo{year}{2018}), \bibinfo{pages}{86--97}.
\newblock


\bibitem[Palomba et~al\mbox{.}(2013)]%
        {palomba2013detecting}
\bibfield{author}{\bibinfo{person}{Fabio Palomba}, \bibinfo{person}{Gabriele
  Bavota}, \bibinfo{person}{Massimiliano Di~Penta}, \bibinfo{person}{Rocco
  Oliveto}, \bibinfo{person}{Andrea De~Lucia}, {and} \bibinfo{person}{Denys
  Poshyvanyk}.} \bibinfo{year}{2013}\natexlab{}.
\newblock \showarticletitle{Detecting bad smells in source code using change
  history information}. In \bibinfo{booktitle}{\emph{2013 28th IEEE/ACM
  International Conference on Automated Software Engineering (ASE)}}. IEEE,
  \bibinfo{pages}{268--278}.
\newblock


\bibitem[Rafighi et~al\mbox{.}(2015)]%
        {rafighi2015studying}
\bibfield{author}{\bibinfo{person}{Masoud Rafighi}, \bibinfo{person}{Yaghoub
  Farjami}, {and} \bibinfo{person}{Nasser Modiri}.}
  \bibinfo{year}{2015}\natexlab{}.
\newblock \showarticletitle{Studying the deficiencies and problems of different
  architecture in developing distributed systems and analyze the existing
  solution}. In \bibinfo{booktitle}{\emph{2015 2nd International Conference on
  Knowledge-Based Engineering and Innovation (KBEI)}}. IEEE,
  \bibinfo{pages}{826--834}.
\newblock


\bibitem[Ramos(2016)]%
        {ramos2016analise}
\bibfield{author}{\bibinfo{person}{Cesar~Coutinho Ramos}.}
  \bibinfo{year}{2016}\natexlab{}.
\newblock \showarticletitle{An{\'a}lise e aplica{\c{c}}{\~a}o de m{\'e}todos de
  modulariza{\c{c}}{\~a}o no desenvolvimento do produto}.
\newblock  (\bibinfo{year}{2016}).
\newblock


\bibitem[Richardson(2018)]%
        {richardson2018microservices}
\bibfield{author}{\bibinfo{person}{Chris Richardson}.}
  \bibinfo{year}{2018}\natexlab{}.
\newblock \bibinfo{booktitle}{\emph{Microservices patterns}}.
\newblock \bibinfo{publisher}{Manning Publications Company,}.
\newblock


\bibitem[Rocha(2018)]%
        {rocha2018monolise}
\bibfield{author}{\bibinfo{person}{Diego Pereira~da Rocha}.}
  \bibinfo{year}{2018}\natexlab{}.
\newblock \bibinfo{title}{Mon{\'o}lise: Uma t{\'e}cnica para
  decomposi{\c{c}}{\~a}o de aplica{\c{c}}{\~o}es monol{\'\i}ticas em
  microsservi{\c{c}}os}.
\newblock
\newblock


\bibitem[Rubert and Farias(2021)]%
        {rubert2021effects}
\bibfield{author}{\bibinfo{person}{Maluane Rubert} {and}
  \bibinfo{person}{Kleinner Farias}.} \bibinfo{year}{2021}\natexlab{}.
\newblock \showarticletitle{On the Effects of Continuous Delivery on Code
  Quality: A Case Study in Industry}.
\newblock \bibinfo{journal}{\emph{Computer Standards and Interfaces}}
  (\bibinfo{year}{2021}), \bibinfo{pages}{103588}.
\newblock


\bibitem[Rudrabhatla(2020)]%
        {artigo3}
\bibfield{author}{\bibinfo{person}{Chaitanya Rudrabhatla}.}
  \bibinfo{year}{2020}\natexlab{}.
\newblock \showarticletitle{Impacts of Decomposition Techniques on Performance
  and Latency of Microservices}.
\newblock \bibinfo{journal}{\emph{International Journal of Advanced Computer
  Science and Applications}}  \bibinfo{volume}{11} (\bibinfo{date}{01}
  \bibinfo{year}{2020}).
\newblock
\urldef\tempurl%
\url{https://doi.org/10.14569/IJACSA.2020.0110803}
\showDOI{\tempurl}


\bibitem[Sjoberg et~al\mbox{.}(2002)]%
        {sjoberg2002conducting}
\bibfield{author}{\bibinfo{person}{Dag~IK Sjoberg}, \bibinfo{person}{Bente
  Anda}, \bibinfo{person}{Erik Arisholm}, \bibinfo{person}{Tore Dyba},
  \bibinfo{person}{Magne Jorgensen}, \bibinfo{person}{Amela Karahasanovic},
  \bibinfo{person}{Espen~Frimann Koren}, {and} \bibinfo{person}{Marek
  Vok{\'a}c}.} \bibinfo{year}{2002}\natexlab{}.
\newblock \showarticletitle{Conducting realistic experiments in software
  engineering}. In \bibinfo{booktitle}{\emph{Proceedings international
  symposium on empirical software engineering}}. IEEE, \bibinfo{pages}{17--26}.
\newblock


\bibitem[Tapia et~al\mbox{.}(2020)]%
        {artigo4}
\bibfield{author}{\bibinfo{person}{Freddy Tapia},
  \bibinfo{person}{Miguel~{\'A}ngel Mora}, \bibinfo{person}{Walter Fuertes},
  \bibinfo{person}{Hern{\'a}n Aules}, \bibinfo{person}{Edwin Flores}, {and}
  \bibinfo{person}{Theofilos Toulkeridis}.} \bibinfo{year}{2020}\natexlab{}.
\newblock \showarticletitle{From Monolithic Systems to Microservices: A
  Comparative Study of Performance}.
\newblock \bibinfo{journal}{\emph{Applied Sciences}} \bibinfo{volume}{10},
  \bibinfo{number}{17} (\bibinfo{year}{2020}), \bibinfo{pages}{5797}.
\newblock


\bibitem[Thakare et~al\mbox{.}(2012)]%
        {thakare2012software}
\bibfield{author}{\bibinfo{person}{Sheetal Thakare}, \bibinfo{person}{Savita
  Chavan}, {and} \bibinfo{person}{PM Chawan}.} \bibinfo{year}{2012}\natexlab{}.
\newblock \showarticletitle{Software Testing Strategies and Techniques}.
\newblock \bibinfo{journal}{\emph{International Journal of Emerging Technology
  and Advanced Engineering}}  \bibinfo{volume}{2} (\bibinfo{year}{2012}),
  \bibinfo{pages}{980--986}.
\newblock


\bibitem[Ulrich(1994)]%
        {ulrich1994fundamentals}
\bibfield{author}{\bibinfo{person}{Karl Ulrich}.}
  \bibinfo{year}{1994}\natexlab{}.
\newblock \showarticletitle{Fundamentals of product modularity}.
\newblock In \bibinfo{booktitle}{\emph{Management of Design}}.
  \bibinfo{publisher}{Springer}, \bibinfo{pages}{219--231}.
\newblock


\bibitem[Urdangarin et~al\mbox{.}(2021)]%
        {urdangarin2021mon4aware}
\bibfield{author}{\bibinfo{person}{Roger~Gon{\c{c}}alves Urdangarin},
  \bibinfo{person}{Kleinner Farias}, {and} \bibinfo{person}{Jorge Barbosa}.}
  \bibinfo{year}{2021}\natexlab{}.
\newblock \showarticletitle{Mon4Aware: A multi-objective and context-aware
  approach to decompose monolithic applications}. In
  \bibinfo{booktitle}{\emph{XVII Brazilian Symposium on Information Systems}}.
  \bibinfo{pages}{1--9}.
\newblock


\bibitem[Villamizar et~al\mbox{.}(2015)]%
        {artigo2}
\bibfield{author}{\bibinfo{person}{Mario Villamizar}, \bibinfo{person}{Oscar
  Garc{\'e}s}, \bibinfo{person}{Harold Castro}, \bibinfo{person}{Mauricio
  Verano}, \bibinfo{person}{Lorena Salamanca}, \bibinfo{person}{Rubby
  Casallas}, {and} \bibinfo{person}{Santiago Gil}.}
  \bibinfo{year}{2015}\natexlab{}.
\newblock \showarticletitle{Evaluating the monolithic and the microservice
  architecture pattern to deploy web applications in the cloud}. In
  \bibinfo{booktitle}{\emph{2015 10th Computing Colombian Conference (10CCC)}}.
  IEEE, \bibinfo{pages}{583--590}.
\newblock


\bibitem[Wohlin et~al\mbox{.}(2012)]%
        {wohlin2012experimentation}
\bibfield{author}{\bibinfo{person}{Claes Wohlin}, \bibinfo{person}{Per
  Runeson}, \bibinfo{person}{Martin H{\"o}st}, \bibinfo{person}{Magnus~C
  Ohlsson}, \bibinfo{person}{Bj{\"o}rn Regnell}, {and} \bibinfo{person}{Anders
  Wessl{\'e}n}.} \bibinfo{year}{2012}\natexlab{}.
\newblock \bibinfo{booktitle}{\emph{Experimentation in software engineering}}.
\newblock \bibinfo{publisher}{Springer Science \& Business Media}.
\newblock


\end{thebibliography}

\end{document}